\documentclass[11pt,  a4paper, showkeys, superscriptaddress,nofootinbib]{revtex4}
 
\usepackage{graphics,epsfig }

\usepackage{graphicx}
 \usepackage{hyperref}
 \usepackage{epstopdf}
\usepackage{amsmath}
\newcommand{\be}{\begin{equation}}
\newcommand{\ee}{\end{equation}}
\newcommand{\bea}{\begin{eqnarray}}
\newcommand{\eea}{\end{eqnarray}}

\begin{document}

\title{\Large Thermodynamic properties, thermal image and phase transition of  Einstein-Gauss-Bonnet black hole  coupled with nonlinear electrodynamics}

\author {Dharm Veer Singh}
\email{veerdsingh@gmail.com}
\affiliation{Department of Physics, Institute of Applied Science and Humanities, GLA University, Mathura, 281406 India.}

\author{Vinod Kumar Bhardwaj}
\email{dr.vinodbhardwaj@gmail.com}
\affiliation{Department of Mathematics, Institute of Applied Science and Humanities, GLA University, Mathura, 281406 India.}

\author{Sudhaker Upadhyay\footnote{Visiting Associate at  IUCAA, Pune, Maharashtra 411007, India}}
\email{sudhakerupadhyay@gmail.com}

\affiliation{Department of Physics, K. L. S. College, Nawada, Bihar 805110, India}
\affiliation{Department of Physics, Magadh University, Bodh Gaya,
 Bihar  824234, India}
 
\affiliation{School of Physics, Damghan University,     Damghan   3671641167,  Iran}
\begin{abstract} 
We obtain  an exact solution of   $AdS$ black hole solution in Einstein-Gauss-Bonnet (EGB) gravity coupled with nonlinear electrodynamics. It interpolates with the $AdS$ regular black hole and $AdS$ EGB black hole in the absence of the Gauss-Bonnet coupling constant and  both magnetic monopole charge   and deviation parameter, respectively. Based on horizon thermodynamics, we
study the  thermodynamic properties of the obtained solution (e.g. mass, temperature, entropy, heat capacity and free energy).  
The Hawking temperature of the nonsingular black hole gets the maximum value at the point where specific heat diverges
and the second-order phase transition occurs at the same point. We find that the smaller nonsingular black holes are stable due to positive heat capacity
and negative free energy. We explicitly trace the relations between the black hole shadow and the critical behavior of charged EGB $AdS$ regular black
hole in the extended phase space.  
\end{abstract}
\maketitle
 
\section{Overview and motivation}
Black holes, one of the most fascinating objects of nature, were predicted by Einstein's gravity in general relativity (GR). This is because the geometric properties of black holes are described by elegant mathematical equations. The recent (astrophysical) observations and experiments associated with gravitational waves \cite{1,2} and   (supermassive) black hole shadows \cite{3} can guide us to test the different gravity models. Due to the limitations of Einstein's gravity,
modifications to Einstein's gravity have been studied widely in the cosmological context. In this connection,  the modifications in the form of higher orders of the Riemann curvature  (and of
its contractions)  had been discussed for the sake of answer to the problem of the dark energy dominance in the evolution of our universe as well as the hierarchy problem. Lee-Wick gravity theories, on the other hand,  are another example of modified gravity that removes the conflict between unitarity and renormalizability in quantum gravity \cite{l,l1,sd}.

In general, gravity theories with higher derivative curvature terms had consistency issues due to negative norm states  (ghost) in their spectra at the quantum level. 
Lovelock gravity is one of the higher derivative gravity theories that maintain second-order equations of motion for the metric and makes the theory ghost-free \cite{lav}.  Gauss-Bonnet (GB) or Einstein-Gauss-Bonnet (EGB)  gravity is particular Lovelock gravity that gets relevance in higher dimensions \cite{4}.
Interestingly, higher orders of the Riemann curvature corrections appear naturally in the low-energy limit of string
theory \cite{5}. This gave rise to considerable interest in EGB gravity theories (especially, the black-hole solutions of these theories) \cite{6,7}.   The GB term is not dynamical  in $4d$ due to total derivative. It becomes dynamical in $4d$ only  when   GB term  must be coupled to a matter field otherwise dimension should be greater than four. Quite interestingly,  by re-scaling the GB coupling
constant $\alpha \to \alpha/d-4$, the  EGB the black hole solutions   in $4d$ have been studied \cite{gla} where the GB term contributes to the (local) dynamics \cite{gla}. The consistency of field-equation  is rather questioned \cite{me}. 
 The limit $d \to 4$ costs us by either partial breaking  of the diffeomorphism or  leading  to additional degrees of freedom.  

 In Refs. \cite{gl1,gl2}, a consistent EGB gravity  in $4d$  has been  proposed that admits   spatial covariance only but    not  the temporal covariance. Recently,  black hole solution and their thermal properties have been studied for  4D $AdS$ EGB  gravity coupled with Yang-Mills field \cite{Singh:2021xbk}.
Moreover, a black hole solution along with its thermodynamics for $4D$ EGB  gravity \cite{fran},  $4D$ EGB massive gravity is presented \cite{Upadhyay:2022axg}, $4D$ EGB wormhole solutions \cite{Godani:2022jwz,Jusufi:2020yus}, and $4D$ EGB quark star \cite{Tangphati:2021tcy,Pretel:2021czp,Tangphati:2021wng,Panotopoulos:2022dbu}.

{From} the perspective of   GR  and its
extensions, nonlinear electrodynamics   is one of the most
crucial material sources of gravity as  it leads to
many interesting geometries, in particular, regular black holes. The nonlinear electrodynamics is an extension of the Maxwell electrodynamics which came into the picture due to Born and Infeld (BI)  in removing the central singularity of a point charge \cite{be}.   A generalized BI action appears naturally in 
superstrings scenario \cite{8,9} induces interest in nonlinear electrodynamics.   Black holes with nonlinear electrodynamics as a matter source are of great importance in astrophysical observations and, therefore, studied extensively  \cite{10,11, nnn,ab,abg11,abg, 12,13,sabir,dvs1,kumar1,dvs2}.  However,   the first regular black hole was proposed by Bardeen \cite{14}.   In the context of nonlinear electrodynamics, it is shown that a regular black hole only describes the formation of a black hole from an initial vacuum region. The preference for nonlinear electrodynamics over Maxwell's theory is only because, along with the conformal breaking, the infinite electric field and self-energy also appear at the position of a point charge.

After Bekenstein and Hawking's proposal of black hole entropy \cite{16} and black hole radiation \cite{17},  black holes are being considered as a thermal system.
Bekenstein stated that the addition of  black hole entropy and the
the entropy of matter is a non-decreasing function of time. 
The generalized second law of thermodynamics suggests that black
holes and self-gravitating systems comprise a larger thermodynamic space.  This leads to the possibility of phase transitions between black holes and self-gravitating systems.
 Understanding such phase transitions is itself important. Such analysis may provide significantly
insights to quantum theories of gravity \cite{15}.

 The plan of the paper is as follows. In Sec. \ref{s2}, we get an exact   black hole solution
for the EGB gravity coupled with nonlinear electrodynamics  and  negative cosmological constant. In Sec. \ref{s3}, we provide a thermal description of the obtained solution by computing mass, temperature and entropy.
Within   section, we also discuss the local and   global  stability of the system. In Sec. \ref{sec:level6},
we study the $P-v$ criticality and phase diagram of the resulting black hole solution. 
In Sec. \ref{s5}, we outline the relation between phase transition and Shadow radius.
Within section, we discuss the effects of various parameter on black hole shadow. 
The discussion of results is reported in the last section. 
\section{$4D$  regular  EGB Black hole solution   }\label{s2}
  The action of EGB gravity coupled   with  nonlinear electrodynamics   and  negative cosmological constant  can be written as
\begin{eqnarray}
S&=&\frac{1}{2}\int d^{d}x\sqrt{-g}\left[ {  R}-2\Lambda +\frac{\alpha}{D-4} {\cal L_{GB}}- 2 { L} (F) \right],
\label{action}
\end{eqnarray}
where $\alpha$ represents the dimensionless GB coupling constant, and ${R}$ denotes  $D$-dimensional  curvature scalar, the GB Lagrangian density is given by $ \mathcal{L_{GB}}:=R_{abcd }R^{abcd}-4R_{ab }R^{ab }+R^{2}$.
Here, ${ L} (F)$ corresponds to matter Lagrangian  which depends on $F=\frac{1}{4} F_{ab}F^{ab}$. \\
Variation the action (\ref{action}) with respect to $g_{ab}$ and the potential $A_a$, we obtained the following equations of motion:
\begin{eqnarray}
&&R_{ab}-\frac{1}{2}\tilde g_{ab}R+\Lambda g_{ab }+\frac{\alpha}{d-4}H_{ab}=T_{ab}\equiv2\left[\frac{\partial {{L(F)}}}{\partial F}F_{a c}F_{\ b}^{c}-\tilde g_{a b}{{L(F)}}\right],\label{egb2}\\
&& \nabla_{a}\left(\frac{\partial {{L(F)}}}{\partial F}F^{a b}\right)=0\qquad \text{and} \qquad \nabla_{a}(* F^{ab})=0,
\label{fe}
\end{eqnarray}
where  GB tensor is expressed as
\begin{eqnarray}
H_{ab }=-\frac{1 }{2}\left[ 8R^{cd }R_{a c b
d }-4R_{a}^{\ cd e }R_{b cde
}-4RR_{ab }+8R_{ac }R_{\ b }^{c }+ g_{ab }\left( R_{cd ef }R^{cd ef }-4R_{cd }R^{cd }+R^{2}\right) \right].
\end{eqnarray}
 Here, we note  that the GB invariant in  $d=4$
  does not contribute to the field equations. So, let us briefly explain the idea of how GB term contributes to the dynamics of gravitational field by re-scaling the GB coupling constant to $\alpha/(d-4)$.  For maximally symmetric solution, Riemann tensor is given by \cite{gla} $R^{ab}_{\ \ cd} =(\delta^a_{\ c}\delta^b_{\ d} -\delta^a_{\ d}\delta^b_{\ c})\Lambda_{eff}/(d-1)$,  where $\Lambda_{eff}$ is an effective cosmological constant. 
To estimate Gauss-Bonnet contribution,  one  calculates
\cite{gla}
\begin{eqnarray}
  \frac{g_{bc}}{\sqrt{-g}}\frac{ {\cal L_{GB}}}{\delta g_{ac}}= \frac{(d-2)(d-3)(d-4) }{2(d-1)}\Lambda_{eff}\delta^a_b.
\end{eqnarray}
Here, the appeared vanishing factor $(d-4)$ 
due to Gauss-Bonnet term in $4D$ gets
canceled by the divergent factor $1/(d-4)$ coming from
the re-scaling GB coupling constant $\alpha/(d-4)$.
 
 Here, we are interested in a particular form of the nonlinear electrodynamics  Lagrangian $L(F )$ which admits the freedom of the duality rotations as discussed in Refs. \cite{nnn, ab, abg11}. This is given by
  \begin{equation}
{{L(F)}}= \frac{F e^{-s(2g^2F)^{1/4}}}{(1+\sqrt{2g^2F})^{3/2}}\left(1+\frac{3}{s}\frac{(2g^2F)^{1/4}}{(1+\sqrt{2g^2F)}}\right),
\label{nonl1}
\end{equation}
where $s=|g|/2M$ with  the free parameters  $g$ and $M$ associated with magnetic charge  and mass, respectively.  For weak fields, this nonlinear electrodynamics correspondence to Maxwell theory. 

To obtain a static spherically symmetric black hole solution in $4D$, we  consider the following line element:
\begin{equation}
ds^2 = -f(r)dt^2+\frac{1}{f(r)} dr^2 + r^2 d\Omega_{2}^2,
\label{metric}
\end{equation}
where $f(r)$ is the metric and  $d\Omega_{2}^2=d\theta^2+\sin^2\theta d\phi^2$  is the metric of a $2$-dimensional sphere. 

To determine the metric function, we consider the following magnetic choice for the Maxwell's field-strength tensor $F_{ab}$ \cite{abg,Singh:2022xgi}
 \begin{equation}
 F_{ab}=2\delta^{\theta}_{[a}\delta^{\phi}_{b]}Y(r,\theta).\label{fab}
 \end{equation}
 Here, we must stress that there exists spherically symmetric solutions with a globally regular metric for gravity coupled to nonlinear electrodynamics  with the Lagrangian $L(F)$, possessing a correct weak field limit,  for 
 the magnetic case only  \cite{Bronnikov:2017tnz,Bronnikov:2000vy}. However, the   electric analogs of magnetic solutions can be found with different Lagrangian (using a Legendre transformation in Hamiltonian formalism) in
different ranges of the radial coordinate \cite{Bronnikov:2000vy}.
 
 With   ansatz (\ref{fab}), the integration of  equation (\ref{fe}) give rise to 
 \begin{equation}
F_{ab}=2\delta^{\theta}_{[a}\delta^{\phi}_{b]}g \sin\theta.
\label{emt2}
\end{equation}
Here,   $g$ is constant (independent of  $r$) is confirmed by using exterior derivative of differential 2-form (\ref{fab}).
Now, the field-strength tensor and, therefore, matter Lagrangian can be simplified to
\begin{equation}
F_{\theta\phi}= g\sin\theta, \qquad F=\frac{1}{2}\frac{g^2}{r^4}, \qquad\text{and} \qquad { L(F)}=\frac{g^2}{2r^2}\frac{e^{-sg/r}}{(r^2+g^2)^{3/2}}\left[1+\frac{g}{s}\frac{r}{(r^2+g^2)}\right].
\label{emt1}
\end{equation}
  
 The non-zero components of energy momentum tensor are given by 
\begin{eqnarray}
  T^t_t=T^r_r&=&\frac{g^2}{2r^2}\frac{e^{-sg/r}}{(r^2+g^2)^{3/2}}\left[1+\frac{g}{s}\frac{r}{(r^2+g^2)}\right],\\
 T^{\theta}_{\theta}=T^{\phi}_{\phi}&=&\frac{g^2}{4r^2}\frac{e^{-sg/r}}{(r^2+g^2)^{3/2}}\left[1+\frac{g}{s}\frac{r}{(r^2+g^2)}-\frac{g(3g^2+8r^2)}{sr}\right.\nonumber\\
 &-&\left. \frac{(3g^2+7r^4)}{r^2(r^2+g^2)^{5/2}}+\frac{gs}{r^3(r^2+g^2)^{3/2}}\right].
\end{eqnarray}
The  $(r,r)$ components of  Eq. (\ref{egb2}) in the limit $d\to 4$ gives
\begin{eqnarray}
 r^5-2r^3\alpha(f-1)\frac{df}{dr}+r^4(f-1)+  r^2\alpha (f-1)^2-\Lambda r^2=\frac{g^2}{4r^2}\frac{e^{-sg/r}}{(r^2+g^2)^{3/2}}\left[1+\frac{g}{s}\frac{r}{(r^2+g^2)}\right].
\label{rr}
\end{eqnarray}
The solution of above equation is given by 
\begin{eqnarray}
f_{\pm}(r)=1+ \frac{r^2}{2\alpha}\pm\frac{r^2}{2\alpha}\sqrt{1+ \frac{8M\alpha\, e^{-k/r}}{(r^2+g^2)^{3/2}}-\frac{4\alpha}{l^2} },
\label{sol1}
\end{eqnarray}
where $k\,(=g^2/2M)$ is called as deviation parameter  that measures the deviation from Schwarzschild black hole solution. For $AdS$,  the cosmological constant ($\Lambda$) can be  expressed in terms of length scale $l$ as $\Lambda=-3/l^2$. Here, $M$ is a constant of integration related to the total mass of the black hole.  The obtained black hole solution (\ref{sol1}) is characterized by the mass, deviation parameter, magnetic charge, GB coupling constant ($\alpha$) and cosmological constant. 
   In the absence of deviation parameter, this solution (\ref{sol1}) identifies  to the following  $4D$ EGB Bardeen black hole   \cite{Singh:2020xju}:
\begin{eqnarray}
f_{\pm}(r)=1+ \frac{r^2}{2\alpha}\pm\frac{r^2}{2\alpha}\sqrt{1+ \frac{8M\alpha\, }{(r^2+g^2)^{3/2}}-\frac{4\alpha}{l^2} }.
\label{sol2}
\end{eqnarray}
 In the absence of  magnetic charge,   this solution identifies to following $4D$ EGB regular black hole \cite{Singh:2020mty}:
\begin{eqnarray}
f_{\pm}(r)=1+ \frac{r^2}{2\alpha}\pm\frac{r^2}{2\alpha}\sqrt{1+ \frac{8M\alpha\, e^{-k/r}}{r^3}-\frac{4\alpha}{l^2} }.
\label{sol3}
\end{eqnarray}
In the absence of GB coupling parameter, this identifies to solution obtained in Ref. \cite{abg}
\begin{eqnarray}
f(r)=1+ \frac{2M r^2 e^{-k/r}}{(r^2+g^2)^{3/2}}+\frac{r^2}{l^2}.
\label{sol4}
\end{eqnarray}
Moreover,  our  solution reduces  to $AdS$ EGB black hole  \cite{gla}, Bardeen black hole,   $AdS$ regular black hole  \cite{tzi} and $AdS$ Schwarzschild black hole   in the limit of   $g=k=0$ and $\alpha=0$, respectively.

 The solutions (\ref{sol1})  behaves asymptotically as
\begin{eqnarray}
&&f_{-}=1-\frac{2Me^{-k/r}}{(r^2+g^2)^{3/2}}+\frac{r^2}{l^2}+\frac{r^2}{2\alpha}+{{\cal O}\left(\frac{1}{r^3}\right)},\label{cc}\\
&&f_+= 1+\frac{2M e^{-k/r}}{(r^2+g^2)^{3/2}}-\frac{r^2}{l^2}+\frac{r^2}{2\alpha}+{{\cal O}\left(\frac{1}{r^3}\right)}.
\end{eqnarray}
Here, one can see that the $+ $ve branch of the solutions is not physical as the positive mass term indicates graviton instabilities and, therefore, this solution will be ruled out.
However,   $-$ve branch of the solutions matches with  the physical  $4D $ $AdS $ EGB black hole solution,  so we must stick with it. In the limit $r \to \infty$ ($M = 0$), the  EGB black hole solution (\ref{cc}) becomes asymptotically flat.  

 In the limit  $k = g = 0$,  the solution (\ref{sol1}) reduces to
\begin{eqnarray}
f_{\pm}(r)=1+\frac{r^2}{2\alpha}\left(1\pm\sqrt{1+\frac{8M\alpha}{r^3}-\frac{ 4\alpha}{l^2}}\,\right).
\label{sole1}
\end{eqnarray}
This solution coincides  with the one obtained by Glavan and Lin in Ref. \cite{gla}. This even matches further with $4D$ $AdS$ Schwarzschild  black hole solution in limit $\alpha \to 0$.

  As we know that  $f(r)=0$ determines the   horizons. Since  Eq. (\ref{sol1}) is a transcendental equation and can not be solved analytically. So, to determine the horizons of the black hole, we solve it numerically and the numerical results are tabulated in the TAB. \ref{t1}.

\begin{table}[h]
	\begin{center}
		\begin{tabular}{c c c c |c c c c}
			\hline
			\hline
			\multicolumn{1}{c}{ }&\multicolumn{1}{c}{ $\alpha=0.1$  }&\multicolumn{1}{c}{}&\multicolumn{1}{c|}{ \,\,\,\,\,\, }&\multicolumn{1}{c}{ }&\multicolumn{1}{c}{}&\multicolumn{1}{c}{ $\alpha=0.2$ }&\multicolumn{1}{c}{}\,\,\,\,\,\,\\
			\hline
			$g$   &	  $r_{+}$  &  $r_{-}$  &  $\delta$  &  $g$   & $r_{+}$  &  $r_{-}$  &  $\delta$  \\
			\hline
			0.20  \ \	&  1.77  \ \ & 0.25 \ \ &  1.52 \ \ &  0.20 \ \ &  1.71	\ \ & 0.33 \ \ & 1.38   \\
			
			0.40 \ \	&  1.64 \ \	 &  0.45 \ \  &  1.19  \ \ &  0.40 \ \	& 1.56 \ \	 & 0.57 \ \ & 0.99  \\  
			
			0.63 \ \	&  1.08	\ \ & 1.08 \ \ &  0.00 \ \  & 	0.57 \ \ &  1.11 \ \ & 1.11 \ \  & 0.00  \\  
\hline
			\multicolumn{1}{c}{ }&\multicolumn{1}{c}{ $\alpha=0.1$  }&\multicolumn{1}{c}{}&\multicolumn{1}{c|}{ \,\,\,\,\,\, }&\multicolumn{1}{c}{ }&\multicolumn{1}{c}{}&\multicolumn{1}{c}{ $\alpha=0.2$ }&\multicolumn{1}{c}{}\,\,\,\,\,\,\\
			\hline	
$k$   &	  $r_{+}$  &  $r_{-}$  &  $\delta$  &  $g$   & $r_{+}$  &  $r_{-}$  &  $\delta$  \\
			\hline
			0.10  \ \	&  1.72  \ \ & 0.33 \ \ &  1.39 \ \ &  0.10 \ \ &   1.65	\ \ & 0.44 \ \ & 1.21   \\
			
			0.30 \ \	&  1.44 \ \	 &  0.51 \ \  &  0.93   \ \ &  0.30 \ \	& 1.31 \ \	 & 0.66 \ \ & 0.65  \\  
			
			0.47 \ \	&  0.92	\ \ & 0.92 \ \ &  0.00 \ \  & 0.385 \ \ &  0.97 \ \ & 0.97 \ \  & 0.00  \\  
			\hline
			\hline	
		\end{tabular}
	\end{center}
	\caption{Cauchy horizon ($r_{-}$) and event horizon ($r_{+}$) and their deviation $\delta=r_+-r_-$ for different values of $g$ and $k$ and fixed value of $l$ and $M$.}
	\label{t1}
\end{table}

\begin{figure}[h]
\begin{tabular}{c c c c}
\includegraphics[width=.50\linewidth]{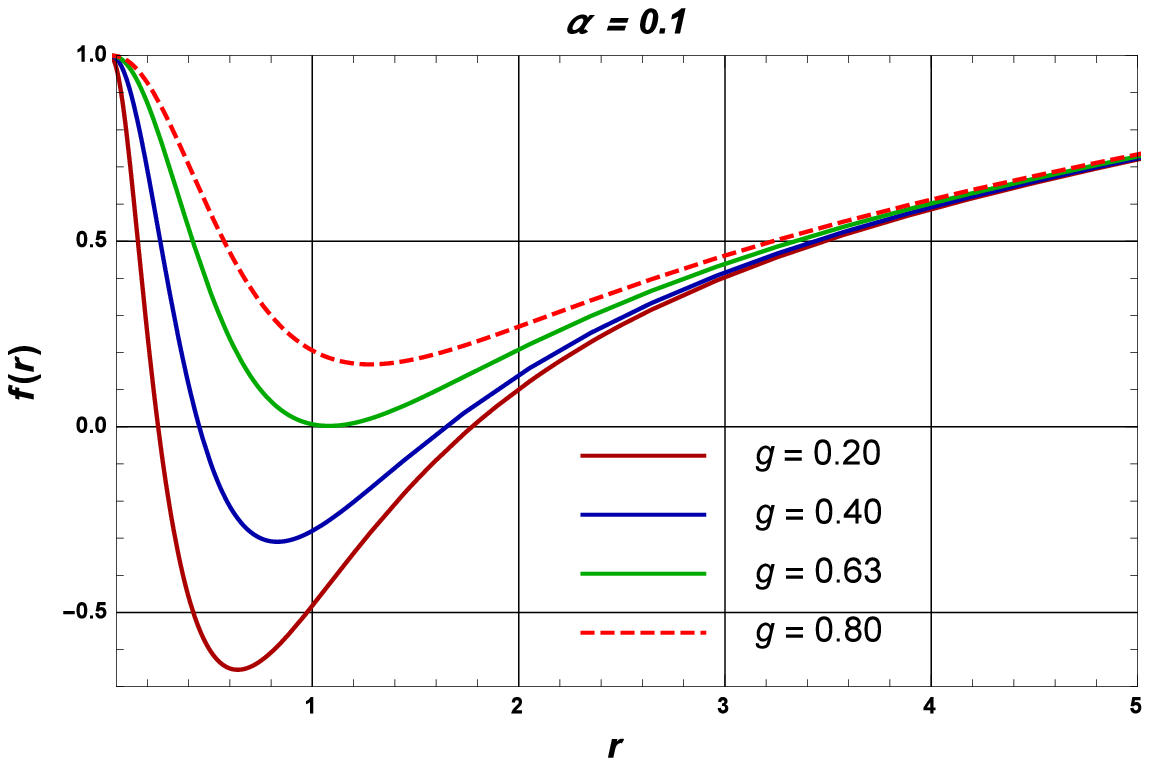}
\includegraphics[width=.50\linewidth]{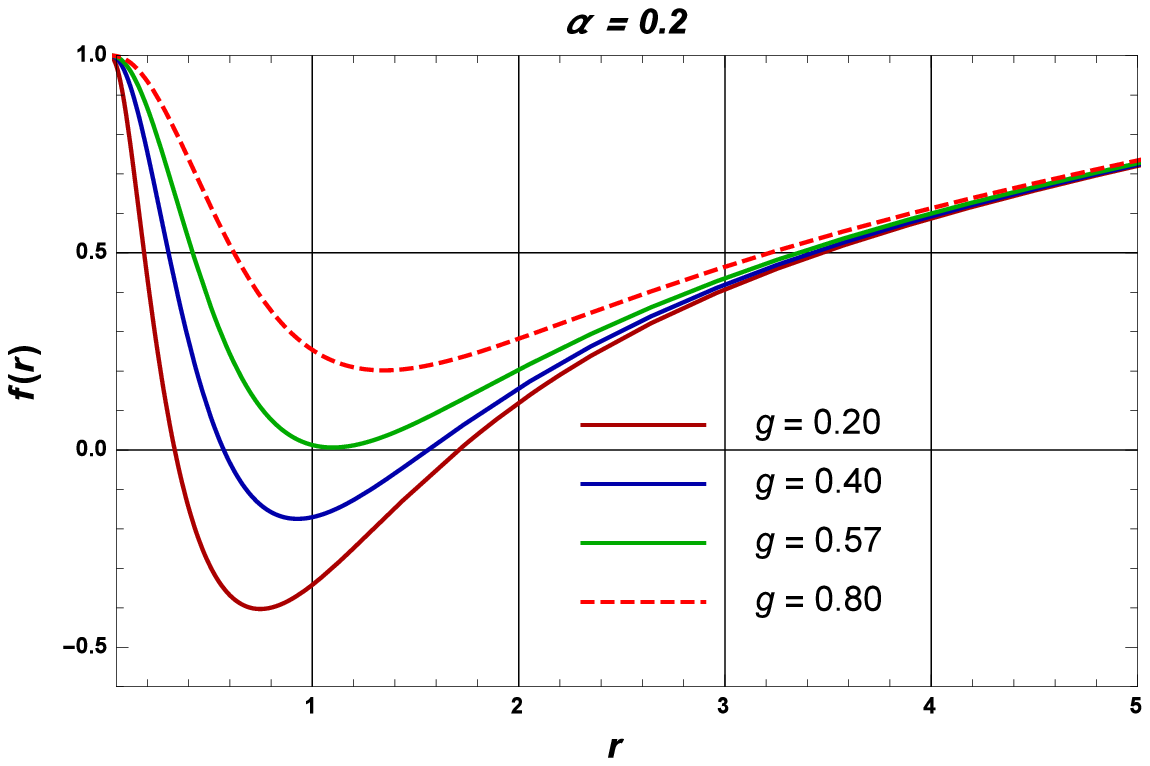} \\
\includegraphics[width=.50\linewidth]{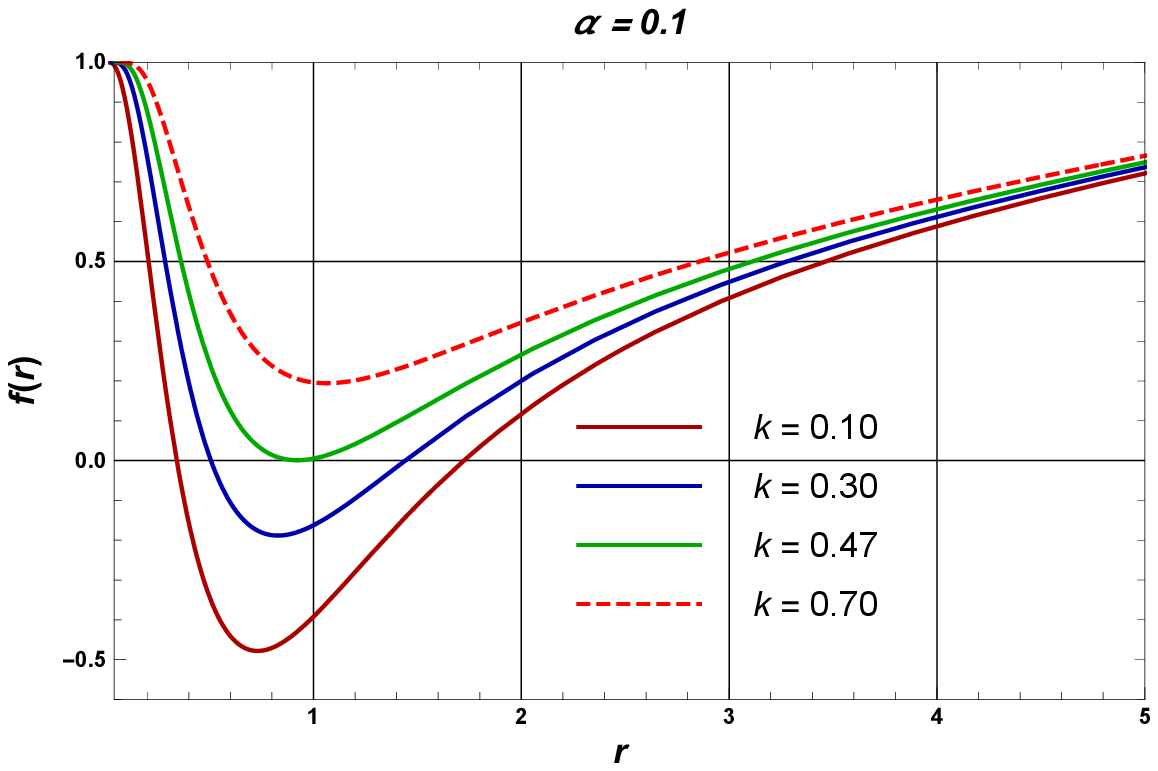}
\includegraphics[width=.50\linewidth]{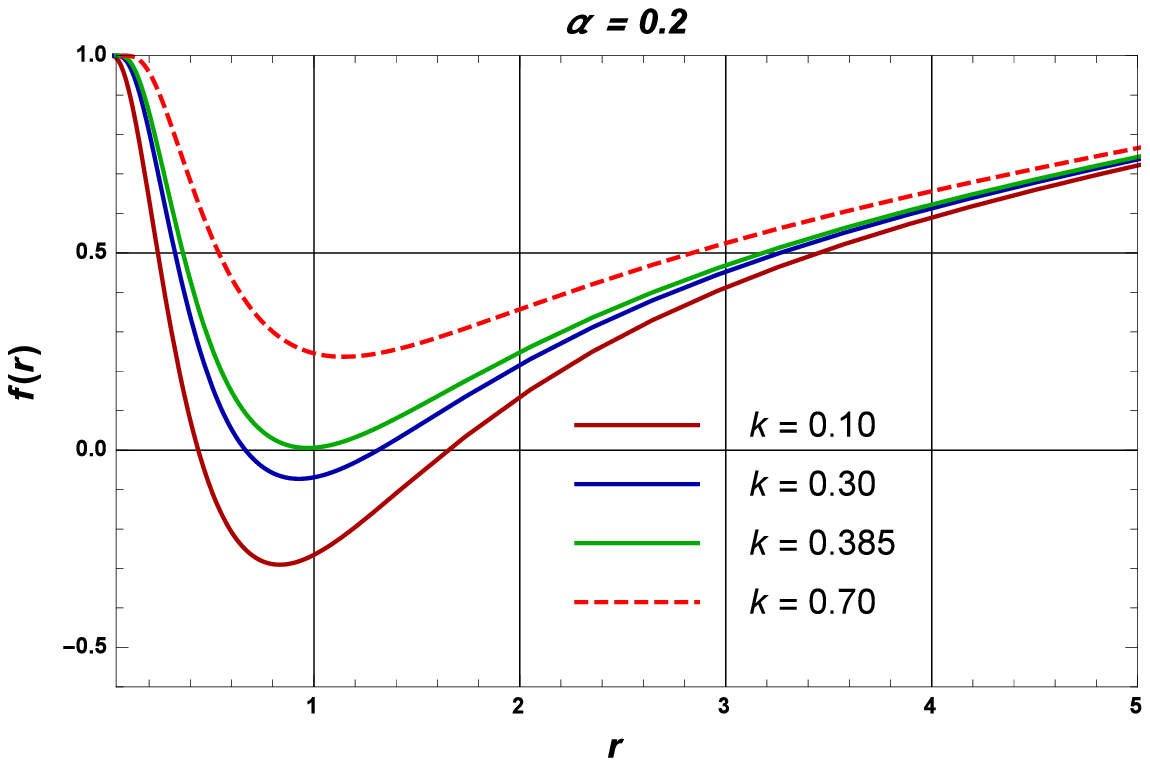}
\end{tabular}
\caption{Metric function   $f(r)$ versus  horizon radius for different  magnetic monopole charge $(g)$ and deviation parameter ($k$) with fixed values of GB parameter $\alpha=0.1$ and $\alpha=0.2$. Here, we set $M=1$. }
\label{fig1} 
\end{figure}
In the Fig. \ref{fig1}, we notice how the horizon structure  depends on   parameters  $k$
\text{and} $g$. The zeros of metric function declares that there is no horizon if  $k > k_c$ (or $g > g_c$), one horizon
 if $k = k_c$ (or $g = g_c$), and are two horizon if
 $k < k_c=0.47$ (or $g < g_c=0.60$). It is worthwhile to mention that the size of horizon increases with decreasing magnetic  charge $(g)$ and deviation parameter $(k)$. The same effect also appears for the GB coupling.
The TABLE \ref{t1} reflects the values of event (outer) horizon and  Cauchy (inner) horizon. 
 
In order to study the nature of singularities of    $AdS$   black hole (\ref{sol1}), curvature invariants  of    Ricci square ($R_{\mu\nu}R^{\mu\nu}$) and Kretshmann scalars ($R_{\mu\nu\lambda\sigma}R^{\mu\nu\lambda\sigma}$) are important.  The  behaviors of the scalar invariants  are given by 
\begin{eqnarray}
&&R=f^{\prime\prime}(r)+\frac{6f^{\prime}(r)}{r}-\frac{6(1-f(r))}{r^2},\\
&&R^{\mu\nu}R_{\mu\nu}=\frac{1}{2}{{f^{\prime
\prime}(r)^2}}+\frac{15}{2}{\left( {\frac{{f^{\prime }(r)}}{{r}}}\right) ^{2}}+\frac{2}{r}f^{\prime\prime}(r)f(r)+\frac{12}{r^4}f(r)^2-\frac{24}{r^4}f(r)\nonumber\\
&&\qquad\qquad-\frac{24}{r^4}(rf^{\prime}(r)+1)+\frac{24}{r^3}f(r)f^{\prime}(r),\\
&&R^{\mu\nu\lambda\rho}R_{\mu\nu\lambda\rho}={{f^{\prime
\prime}(r)^2}}+6{\left( {\frac{{f^{\prime }(r)}}{{r}}}\right) ^{2}}%
+12{\left( {\frac{{f(r)-1}}{{{r^{2}}}}}\right) ^{2}}.
\end{eqnarray}
These invariants are smooth in $r \to 0$ limit and, therefore, signify regular space-time.
  
\section{Thermodynamics}\label{s3}
In this section, we investigate the thermal properties of $4D$ $AdS$ EGB regular black hole.
In this connection, we discuss,  particularly,
   mass, temperature, entropy, specific heat, and Gibbs free energy at the black hole horizon. The mass of the black hole can be estimated from the metric function (\ref{sol1}) on the horizon as
\begin{eqnarray}
 M =\left(g^2+r_{+}^2\right)^{3/2} e^{k/r_{+}} \Big[\frac{(\alpha  + r_{+}^2)}{2 r_{+}^4}+\frac{1}{2 l^2}\Big].
 \label{m}
\end{eqnarray}
This expression represents the mass of regular black hole in $4D$ EGB gravity  (\ref{m}). From the above expression, it is clear that, for $k=0$   it reduces to the mass of the $AdS$ EGB Bardeen  black hole \cite{Singh:2020xju}, for $g=0$ it reduces to $AdS$ EGB regular  black hole \cite{Singh:2020mty} and for both $k=0, g=0$ it reduces to $4D$ EGB black hole \cite{gla}. For vanishing  $\alpha$ and $k$, the above expressing  reduces to the mass of Bardeen black hole \cite{tzi} and for vanishing $\alpha$ and  $g=0$ it reduces to  the mass of $AdS$ regular black hole. Finally, it reduces to mass of the $AdS$ Schwarzschild black hole for vanishing all three parameters:   $g$, $\alpha$ and $k$.

\begin{figure}[h]
	\begin{tabular}{c c c c}
		\includegraphics[width=.50\linewidth]{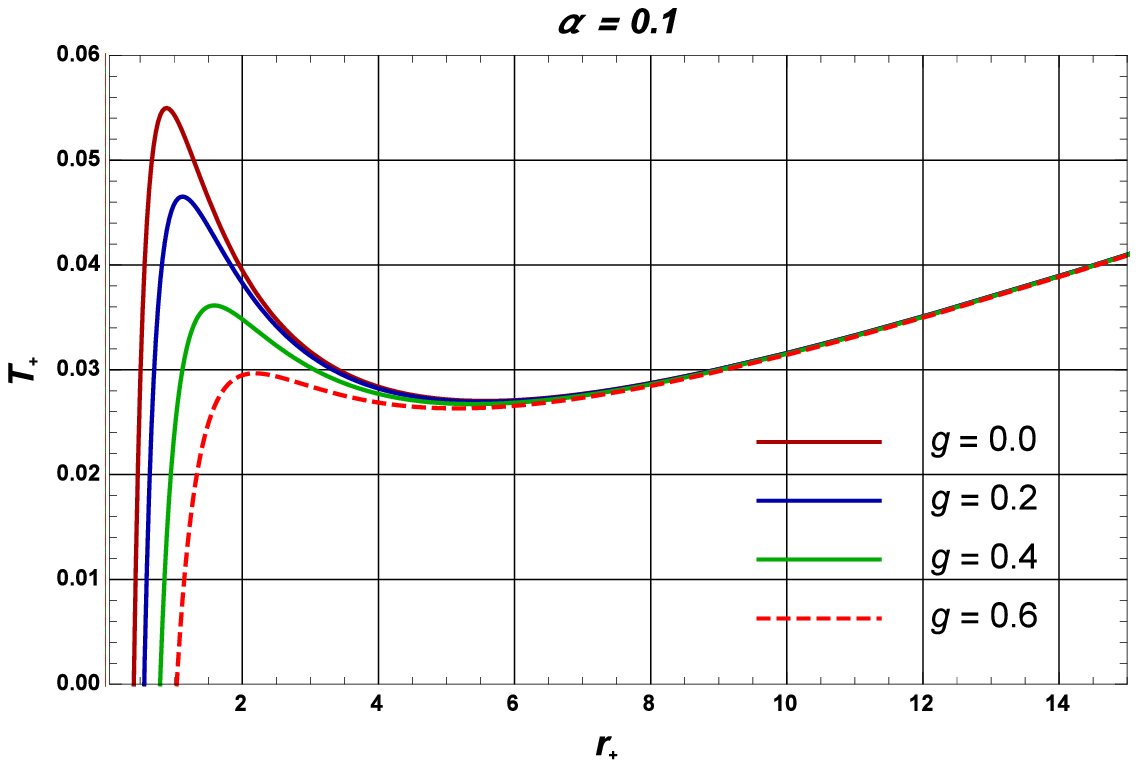}
		\includegraphics[width=.50\linewidth]{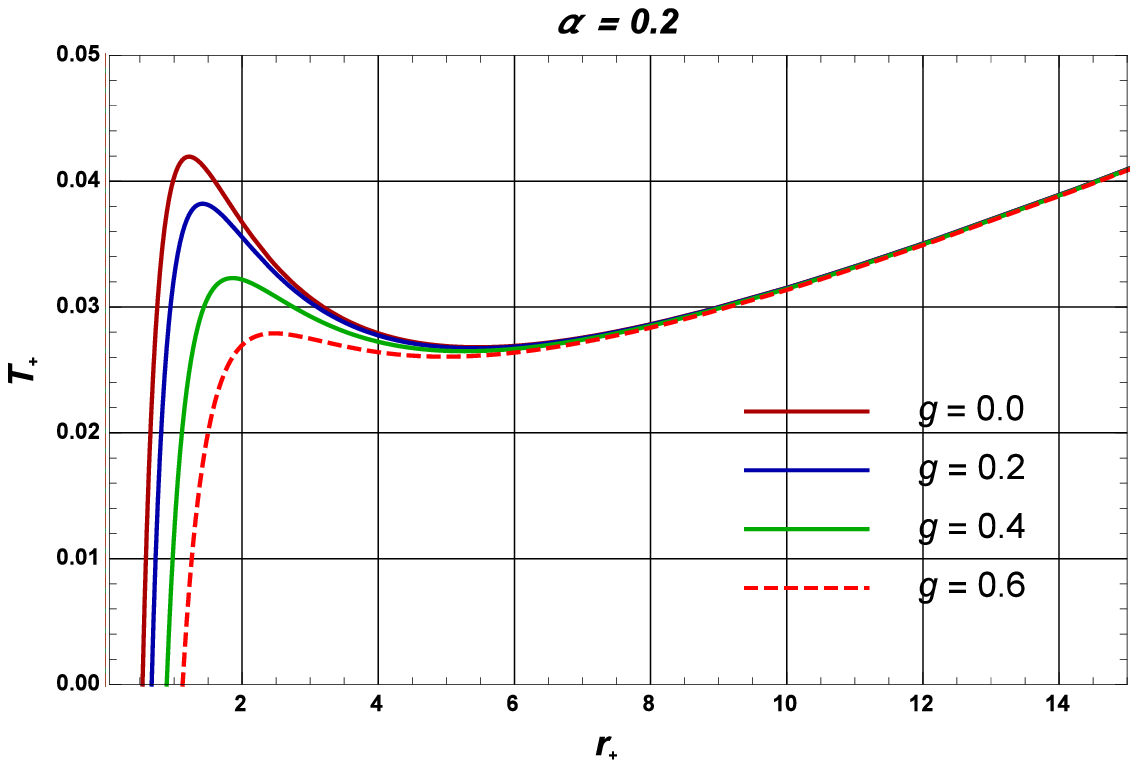} \\
\includegraphics[width=.50\linewidth]{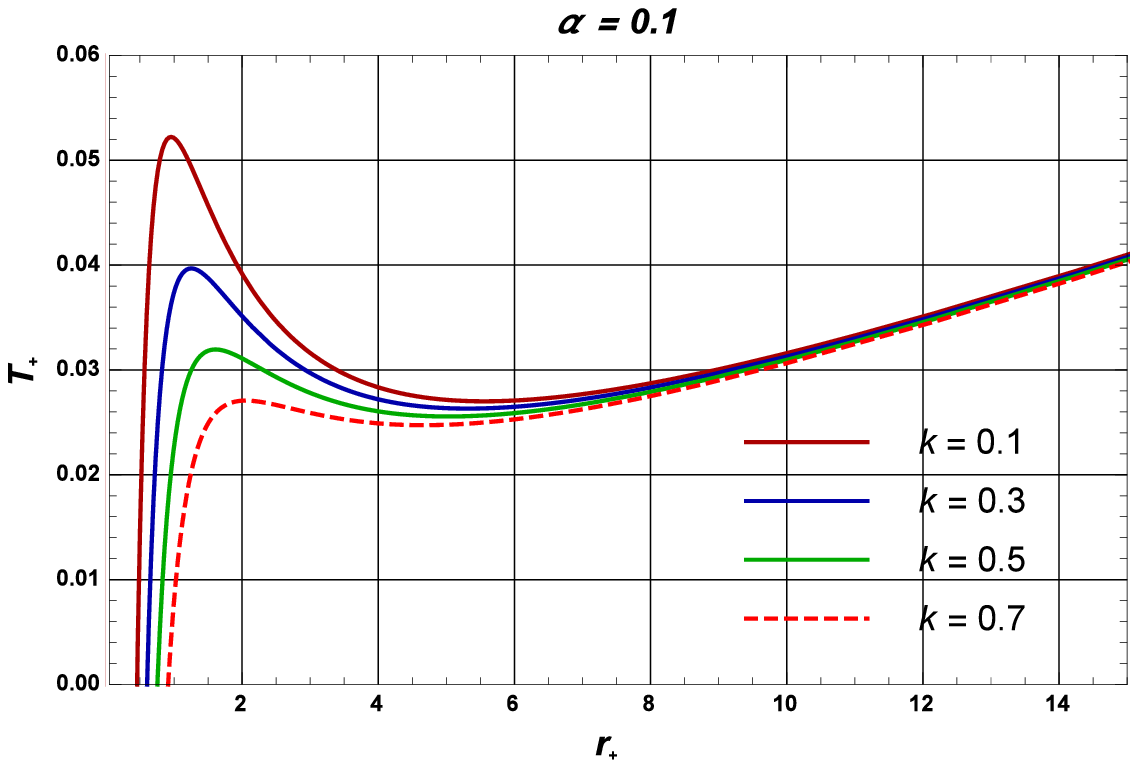}
		\includegraphics[width=.50\linewidth]{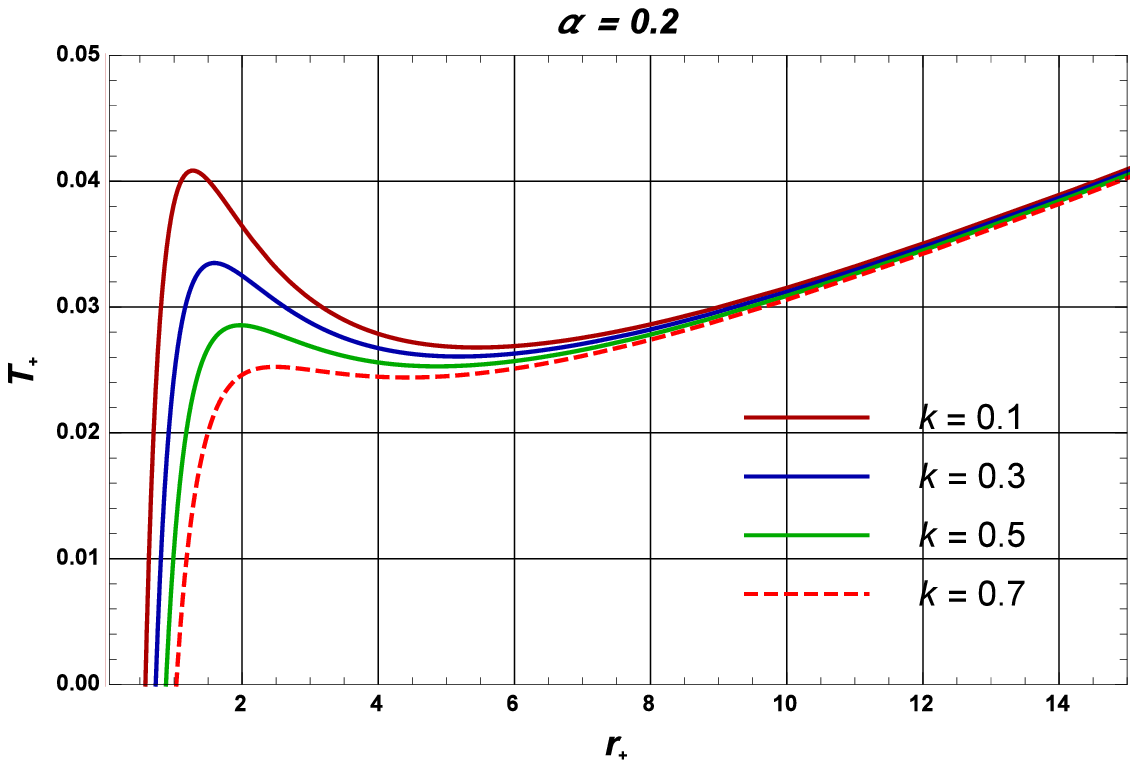} 
	\end{tabular}
	\caption{Temperature versus horizon radius for different valued of  magnetic monopole charge $(g)$ and deviation parameter ($k$) with fixed values of  GB parameter $\alpha=0.1$ and $\alpha=0.2$. Here, we set $ l=10$. }
	\label{fig3} 
\end{figure}

The standard formula to calculate Hawking temperature  is given by
\begin{equation}
T_+=\frac{1}{2\pi}{\sqrt{-\frac{1}{2}\nabla_\mu\xi_\nu \nabla^\mu \xi^\nu}}=\frac{1}{4\pi}f'(r_{+}).
\end{equation} 
In our case, this leads to
\begin{eqnarray}
T_+&=&\frac{1}{4 \pi r_{+}} \bigg(\frac{3 r_{+}^7+l^2 r_{+}^2 (\alpha+r_{+}^2)(9r-2k)-k (g^2+r_{+}^2) (\alpha l^2+l^2 r_{+}^2+r_{+}^4)}{l^2 r_{+} (g^2+r_{+}^2) (r_{+}^2+2\alpha)} \bigg).
\label{t}
\end{eqnarray}
From the expression of  Hawking temperature it is clear that  when $k=0$   it identifies  to the Hawking temperature of the $AdS$ EGB Bardeen  black hole \cite{Singh:2020xju} and   and to the $AdS$ EGB regular  black hole \cite{Singh:2020mty} for $g=0$. In the limit of $\alpha =0$, the Hawking temperature reduces to
\begin{eqnarray}
T_+&=&\frac{1}{4 \pi r_{+}} \bigg(\frac{3 r_{+}^7+l^2 r_{+}^4(9r-2k)-k (g^2+r_{+}^2) (  l^2 r_{+}^2+r_{+}^4)}{l^2 r^3_{+} (g^2+r_{+}^2) (r_{+}^2+2)} \bigg).
\label{t10}
\end{eqnarray} 
 In order to analyze the behavior of temperature for  the various values of magnetic charge $g$ and the deviation parameter $k$, we plot  Fig. \ref{fig3}.
 Here, we see that the temperature first grows to a maximum value of $T_+^{max}$ and then it drops to a minimum value, and then it starts growing again. 

The maximum values of temperature for different values of parameters are tabulated in TABLES \ref{f4} and \ref{f5}.
\begin{table}[h]
	\begin{center}
		\begin{tabular}{c c  c | c c c}
			\hline
			\hline
			\multicolumn{1}{c}{ }&\multicolumn{1}{c}{$\alpha=0.1$, \ $k=0.1$  }&\multicolumn{1}{c|}{ \,\,\,\,\,\, }&\multicolumn{1}{c}{}&\multicolumn{1}{c}{$\alpha=0.2$, \ $k=0.1$}&\multicolumn{1}{c}{ }\,\,\,\,\,\,\\
			\hline
			 $g$  & $r$   &  $T_+^{max}$  &  $g$   & $r$  &  $T_+^{max}$  \\
			\hline
			0.00  \ \	&  0.89  \ \ & 	0.055 \ \ & \ 0.00 \ \ &  1.22 \ \ & 0.042   \\
			
			0.20 \ \ &  1.12 \ \	 & 	0.047  \ \ & \ 0.20  \ \	& 1.42 \ \ & 0.038  \\  
			
			0.40 \ \	&  1.59	\ \ & 0.036  \ \ & \ 0.40 \ \ &  1.86 \ \   & 0.032  \\  
			
			\hline
			\hline	
		\end{tabular}
	\end{center}
	\caption{The values of maximum Hawking temperature for the $4D$  AdS EGB regular black hole with different values of parameters and fixed value of $l=10$.}
	\label{f4}
\end{table}	
\begin{table}[ht]
	\begin{center}
			\begin{tabular}{c c  c | c c c}
			\hline
			\hline
			\multicolumn{1}{c}{ }&\multicolumn{1}{c}{$\alpha=0.1$, \ $g=0.1$  }&\multicolumn{1}{c|}{ \,\,\,\,\,\, }&\multicolumn{1}{c}{}&\multicolumn{1}{c}{$\alpha=0.2$, \ $g=0.1$}&\multicolumn{1}{c}{ }\,\,\,\,\,\,\\
			\hline
			 $k$  & $r$   &  $T_+^{max}$  &  $k$   & $r$  &  $T_+^{max}$  \\
			\hline
		0.10  \ \	&  0.95  \ \ & 0.052  \ \ & \ 0.10  \ \ &  1.27 \ \ &  0.040 \\
			
		 0.30 \ \	&  1.25 \ \	 & 0.039   \ \ & \ 0.30  \ \	& 1.62 \ \ &  0.035 \\  
			
		0.50 \ \	&  1.60	\ \ &  0.032 \ \  & \ 0.50 \ \ &  2.00 \ \   & 0.029  \\  
			
			\hline
			\hline	
		\end{tabular}
	\end{center}
	\caption{The values of maximum Hawking temperature for the $4D$  AdS EGB regular black hole with different values of parameters and fixed value of $l=10$.}
	\label{f5}
\end{table}
 From these TABLES, we see that the values of maximum temperature decrease with an increase in the values of $g$ and $k$. The maximum temperature diverges when the horizon radius shrinks to zero.  Fig. \ref{fig3} also suggests that the Hawking temperature vanishes when the two horizons  (Cauchy and event) merge.

\noindent The thermodynamical quantities must follow  the first law of thermodynamics
\be
dM =T_{+}dS_{+}.
\ee
Since we have explicit expressions of mass and Hawking temperature, then the first law 
gives the value of entropy as 
\begin{eqnarray}
S_+=\pi r_+^2 +4\alpha \log [r_{+}].
\end{eqnarray}
Now, we are interested in the stability of this black hole. We are aware that local stability is decided by the signature of heat capacity. The positive indicates that the system is stable 
  and the negative indicates the unstable one  \cite{sabir,dvs1,kumar1}.  
  However, the global stability can be studied from the Gibbs free energy. 
  
Now, we first calculate from the given formula
\be
C_+=\frac{\partial M }{\partial T_+}=\bigg(\frac{\partial M }{\partial r_+}\bigg)\bigg(\frac{\partial r_+}{\partial T_+}\bigg).
\label{sh1}
\ee
 This gives
\be
C_+=-\frac{2 \pi  \left(g^2+r_{+}^2\right)^{5/2} e^{k/r_{+}} \left(2 \alpha +r_{+}^2\right)^2 \bigg(-k \left(g^2+r_{+}^2\right) \left(l^2 \left(\alpha +r_{+}^2\right)+r_{+}^4\right)+\mathcal{X}\bigg)}{r_{+}^3 \bigg(2 l^2 g^4  r_{+} \left(2 \alpha +r_{+}^2\right)^2 +2 k \left(g^2+r_{+}^2\right)^2 \mathcal{W} +\mathcal{Y}+\mathcal{Z}\bigg)},
\label{sh2}
\ee
where 
\begin{eqnarray}
&&\mathcal{W}=  l^2  \left(2 \alpha^2 +r_{+}^4+2 \alpha  r_{+}^2\right)-2\alpha r_{+}^4,\\ 
&& \mathcal{X} = -2 g^2 l^2 r_{+} \left(2 \alpha +r_{+}^2\right)+l^2 r_{+}^3 \left(r_{+}^2-\alpha \right)+3 r_{+}^7,\\ 
&&\mathcal{Y} = g^2 r_{+}^2 \left(l^2 \left(7 r_{+}^5+31 \alpha  r_{+}^3+22 \alpha^2 r_{+} \right)+9 r_{+}^7+30 \alpha  r_{+}^5\right),\\
&&\mathcal{Z} = r_{+}^4 \left(l^2 \left(-r_{+}^5+5 \alpha  r_{+}^3+2 \alpha^2 r_{+} \right)+3 \left(r_{+}^7+6 \alpha  r_{+}^5\right)\right).
\end{eqnarray}

From the Eq. (\ref{sh2}), it is evident that the heat capacity depends on all the 
parameters   $g$,    $k$, $\alpha$, and   $\Lambda$. For $k=0$, it coincides with the mass of the heat capacity of $AdS$ EGB Bardeen black hole \cite{Singh:2020xju}. The heat capacity in absence of the GB parameter reduces to
\begin{eqnarray}
C_+=-\frac{2 \pi  \left(g^2+r_{+}^2\right)^{5/2} e^{k/r_{+}} r_{+}^4 \bigg(-k \left(g^2+r_{+}^2\right) \left(l^2 \left(r_{+}^2\right)+r_{+}^4\right)+\mathcal{X}\bigg)}{r_{+}^3 \bigg(2 l^2 g^4  r_{+} \left(r_{+}^2\right)^2 +2 k \left(g^2+r_{+}^2\right)^2 \mathcal{W} +\mathcal{Y}+\mathcal{Z}\bigg)},
\label{spe}
\end{eqnarray}
with
\begin{eqnarray}
&&\mathcal{W} = l^2   r_{+}^4,\qquad \qquad \mathcal{X} = \left(-2 g^2 l^2 r_{+}^3 +l^2 r_{+}^5+3 r_{+}^7\right),\\ 
&&\mathcal{Y} = g^2 r_{+}^2 \left(  7 r_{+}^5l^2  +9 r_{+}^7\right),\qquad\text{and}\qquad \mathcal{Z} = r_{+}^4 \left(-l^2r_{+}^5+3 r_{+}^7\right).
\end{eqnarray}
Here, we observe that the expression of heat capacity we obtain is the most general one.
The various limiting case can be studied just by switching various parameters off. 

\begin{figure}[h]
	\begin{tabular}{c c c c}
		\includegraphics[width=.50\linewidth]{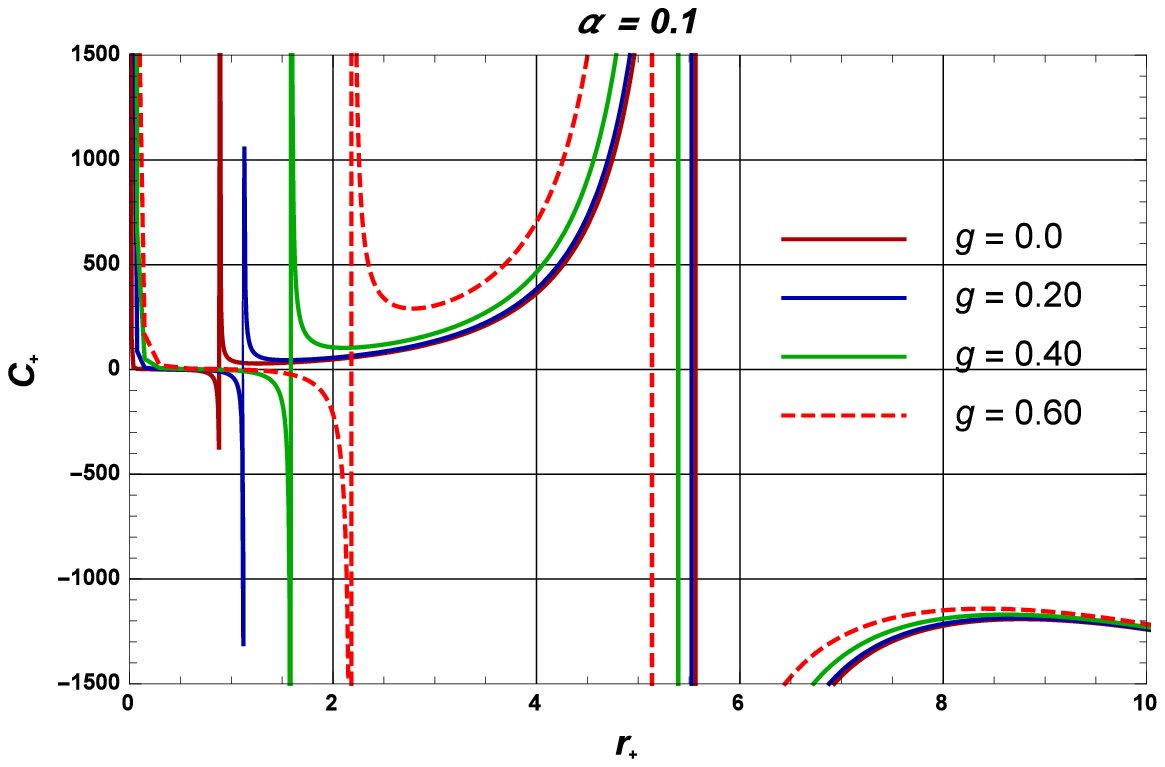}
		\includegraphics[width=.50\linewidth]{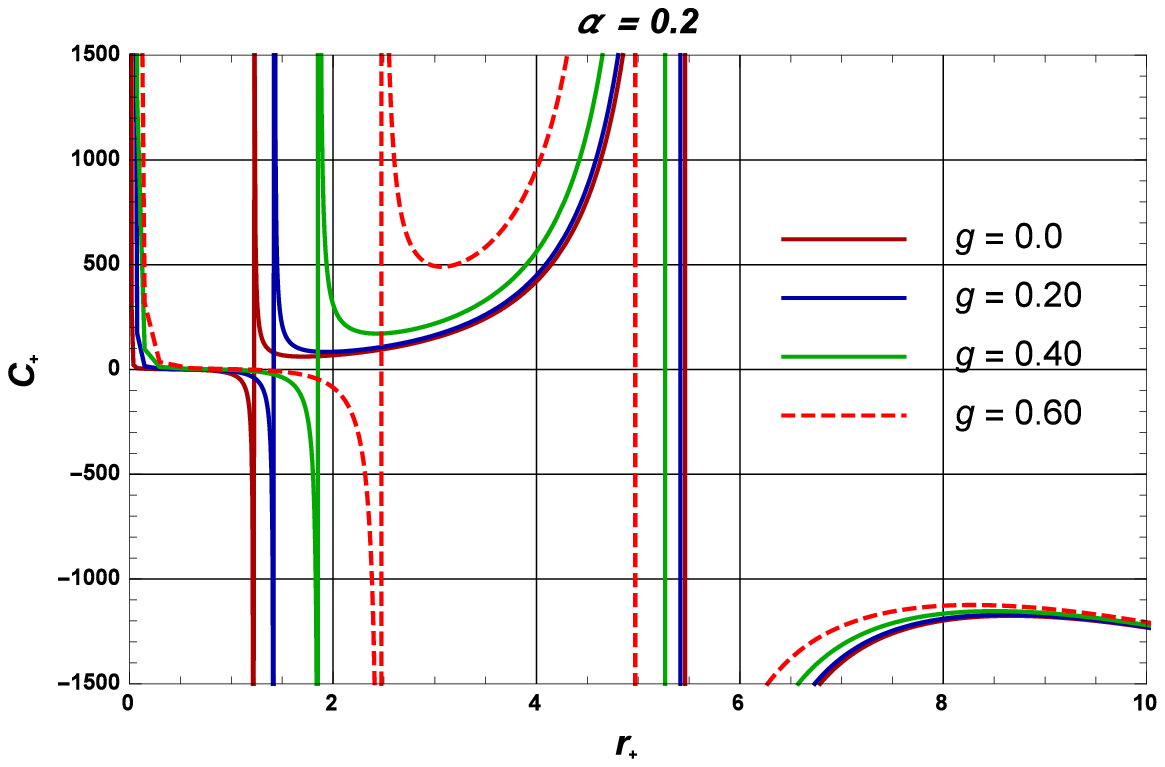} \\
\includegraphics[width=.50\linewidth]{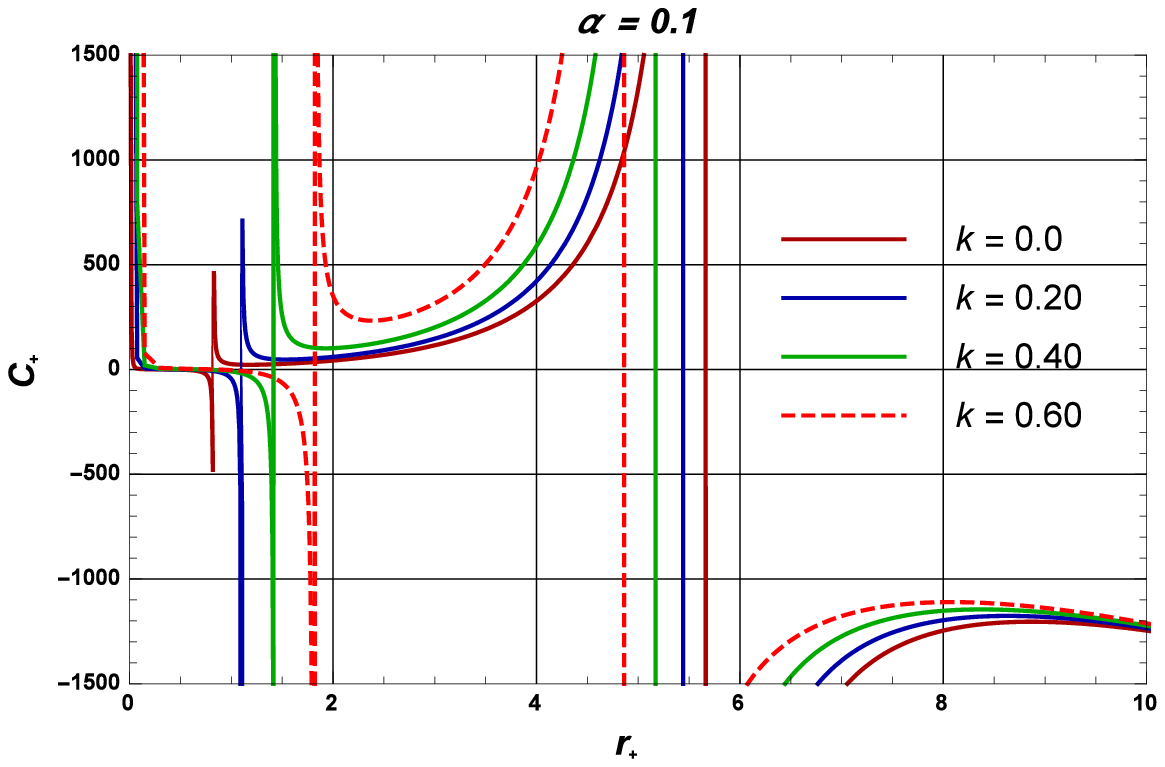}
		\includegraphics[width=.50\linewidth]{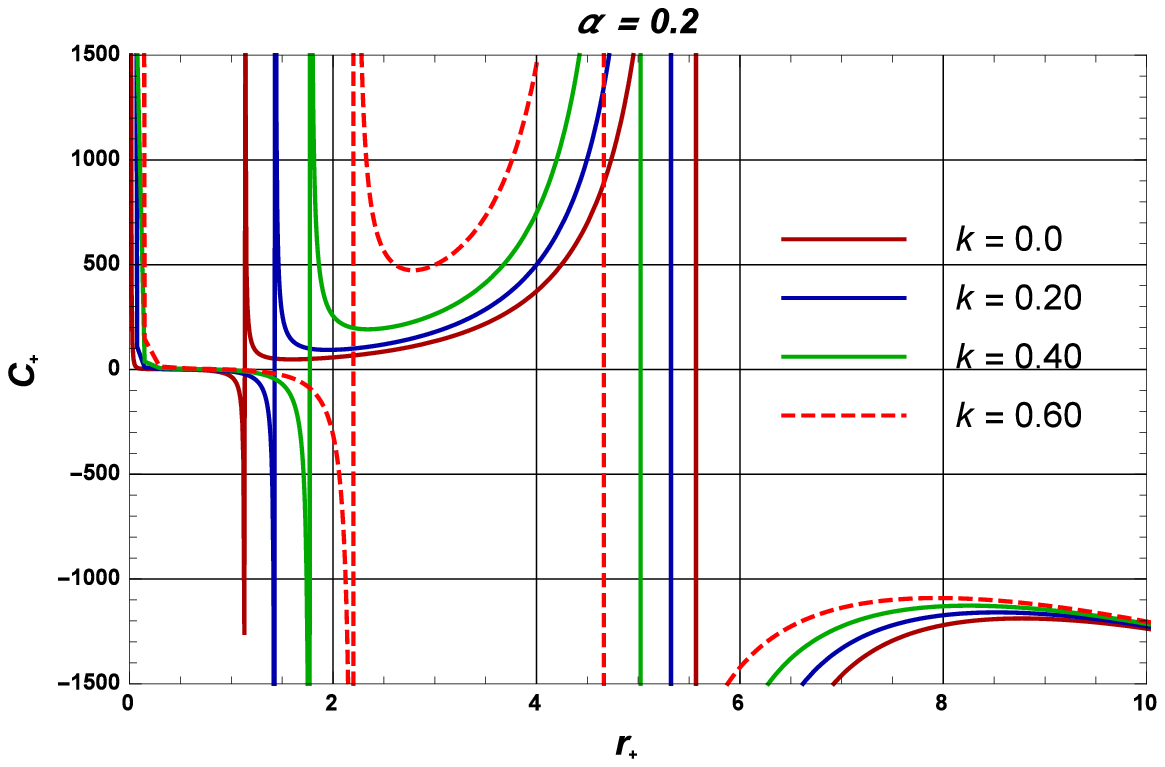} 
	\end{tabular}
	\caption{The  heat capacity  $C_+$ versus horizon radius with the different values of  $g$ and $k$ for two fixed value of GB parameter $\alpha=0.1$ and $\alpha=0.2$.  }
	\label{fig5} 
\end{figure}

Now, the stability can be checked  when one observes the plots of heat capacity for fixing the parameters as depicted in FIG. \ref{fig5}. 
 The heat capacity is showing discontinuity at a particular point (say critical radius $r_+=r_C$) which signifies a second-order phase transition.
 Remarkably, at this point of critical radius,  the temperature attains the maximum value $T_+^{max}$. Hence, a phase transition occurs in-between stable and unstable phases when the black hole gets bigger.

\begin{figure}[h]
	\begin{tabular}{c c c c}
		\includegraphics[width=.50\linewidth]{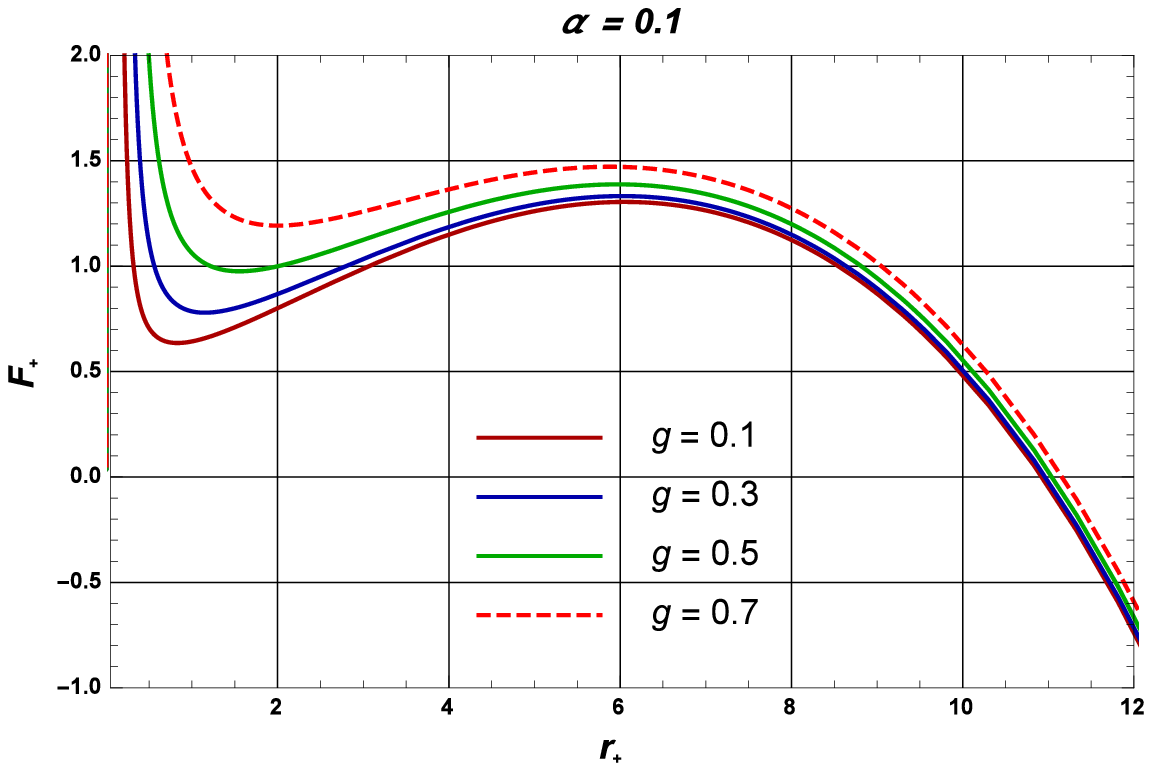}
		\includegraphics[width=.50\linewidth]{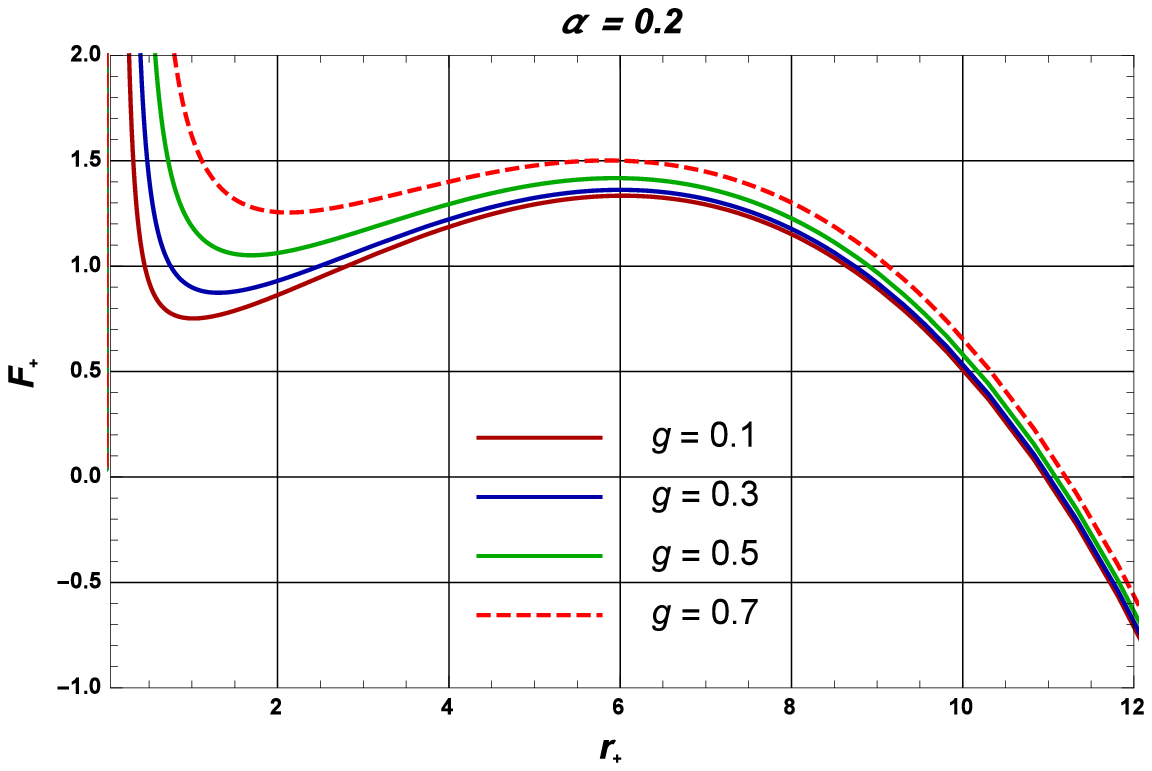} \\
\includegraphics[width=.50\linewidth]{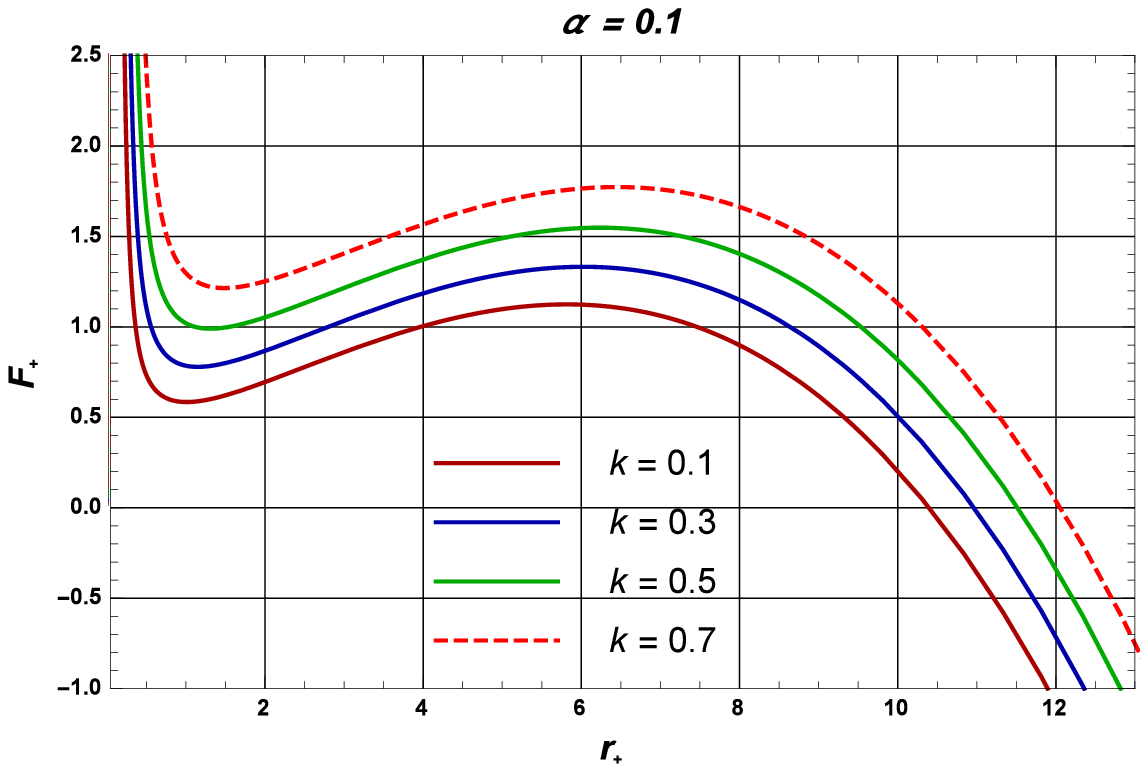}
		\includegraphics[width=.50\linewidth]{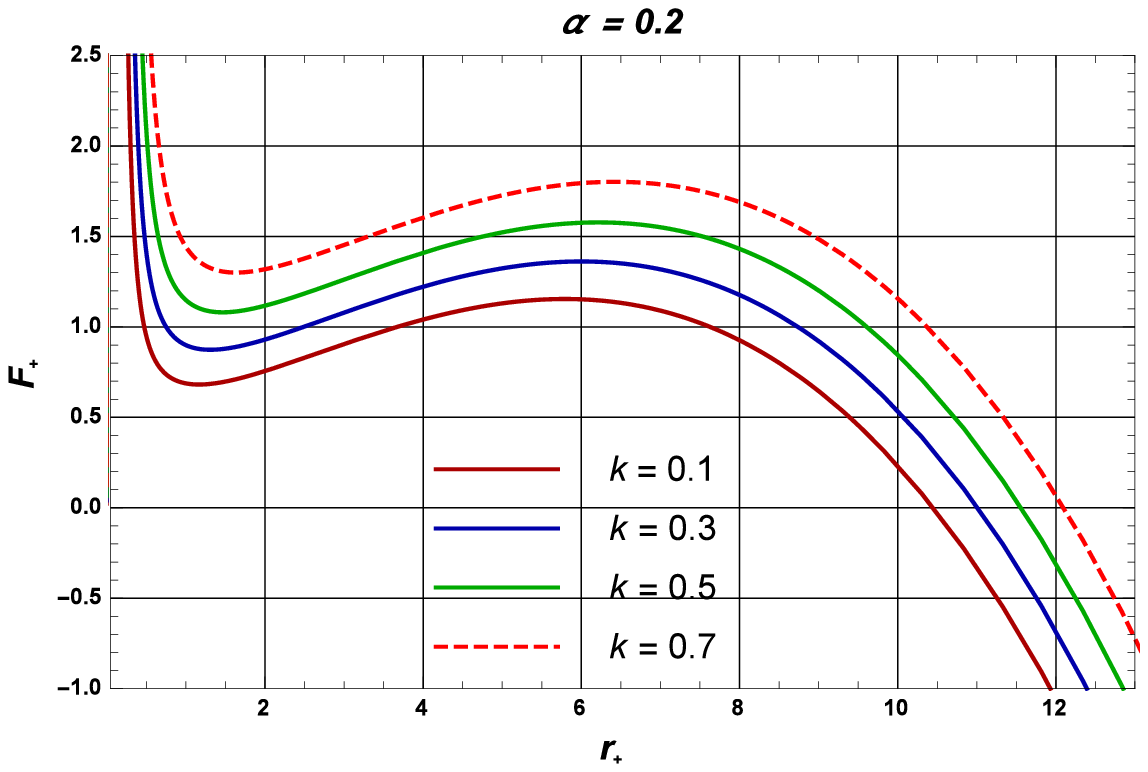} 
	\end{tabular}
	\caption{ Gibbs free energy  versus  horizon radius for different values of   $g$ and $k$ with fixed values of GB parameter $\alpha=0.1$ and $\alpha=0.2$.   }
	\label{fig7} 
\end{figure}

Now we turn to the global stability of the black hole,  which is characterized by Gibbs free energy. The Gibbs free energy can be estimated by standard definition $F_+=M-T_+S_+$, as follows
\bea
F_+&=&\frac{\left(g^2+r_{+}^2\right)^{3/2} \left(\alpha  l^2+l^2 r_{+}^2+r_{+}^4\right) e^{k/r_{+}}}{2 l^2 r_{+}^4}\nonumber\\
&+& \frac{k (g^2+r_{+}^2)(\alpha l^2+l^2 r_{+}^2+r_{+}^4)-3 r_{+}^7-l^2 r_{+}^2 (\alpha+r_{+}^2)(9r_{+}-2k) }{4 l^2 (g^2+r_{+}^2) (r_{+}^2+2\alpha)}.
\label{gibbs}
\eea
Since,  global    stability of the black hole is confirmed by  condition $ F_+\leq 0 $.
Now, to analyze the behavior of this expression, we plot FIG. \ref{fig7}.
  Here, we observe that there exists a  local minimum ($r_{min}$) and maximum ($r_{max}$) which are in agreement with the extremal points of Hawking temperature.  At this point free energy change nature.  After this point ${r_{min}}$, the free energy increases with  horizon radius $r$ and  attains the maximum value at $r_{max}$. After this point, the free energy decreases with the horizon radius.

\section{ $P-v$ Criticality and Phase Diagrams}\label{sec:level6}
Now, we plan to explore the  critical behavior of the present black hole solution in
an extended phase space, by considering $\Lambda$ as the thermodynamic pressure.  The  cosmological constant appears in the pressure as  $P=-\frac{\Lambda}{8\pi}=\frac{3}{8\pi l^2}$.   Plugging the value of $l$ from 
  Eq. (\ref{t}), the form of pressure becomes
\begin{eqnarray}
P = \frac{3T(r_+^6+2r_+^4\alpha+g^2r_+^2(r_+^2+2\alpha))}{2r_+^4(g^2k+(k-3r_+)r_+^2)}+\frac{3(r_+^5-r_+^2+\alpha+(r_+^2+\alpha)[k(r_+^2+g^2)+2g^2r_+])}{8\pi r_+^4(g^2k+(k-3r_+)r_+^2)}.
\label{pv4}
\end{eqnarray}
  The enthalpy in the case of the black hole system is described by the total mass of the system.  
  Therefore, the  thermodynamic volume can be calculated as
\begin{eqnarray}
 V=\left(\frac{\partial M}{\partial P }\right)_{S_+}= \frac{4\pi r_+^3}{3}.
\label{pv3}
\end{eqnarray}
The critical temperature and critical  pressure can be determined from the conditions:
\begin{equation}
\left.\frac{\partial P_+}{\partial r_+}\right|_{T_+}= \left.\frac{\partial ^2P_+}{\partial r_+^2}\right|_{T_+}=0.
\label{pv5}
\end{equation}
For the given pressure (\ref{pv4}), these conditions lead to the following equation:
\begin{eqnarray}
&&r^7(45g^4k+36g^4 r+30 kg^2r^2+45g^2 r^3+9kr^4-3r^5)+r^3\alpha(4g^6k+156r^4kr^2+204g^4r^3+\nonumber\\&&84g^2kr^4+360g^2r^5)+\alpha^2(132kg^4r^2-16g^2k^2-24g^6kr-48g^4k^2r^2+360g^4r^4-48g^2k^2r^4+\nonumber\\&&144kg^2r^5+588g^2r^6-16k^2r^6+84kr^7+36r^8)=0.
\label{pv6}
\end{eqnarray}
The critical points and the horizon radius can be determined by solving the above equation. 
However, one can not solve this equation analytically but can determine the critical radius $r_C$, critical pressure $P_C$, and temperature  $T_C$   numerically. We write the numerical results in  TABLE  \ref{t11} for various values of deviation parameter and magnetic charge. From the table, we find that the value of critical radius increases along with magnetic charge $g$. However, the value of the critical radius decreases along with the deviation parameter $k$. 
Interestingly, the universal ratio $\frac{ P_Cr_C}{T_C}$  increases along with both the magnetic charge $g$ and deviation parameter $k$. 

\begin{figure}[h]
	\begin{tabular}{c c c c}
		\includegraphics[width=.450\linewidth]{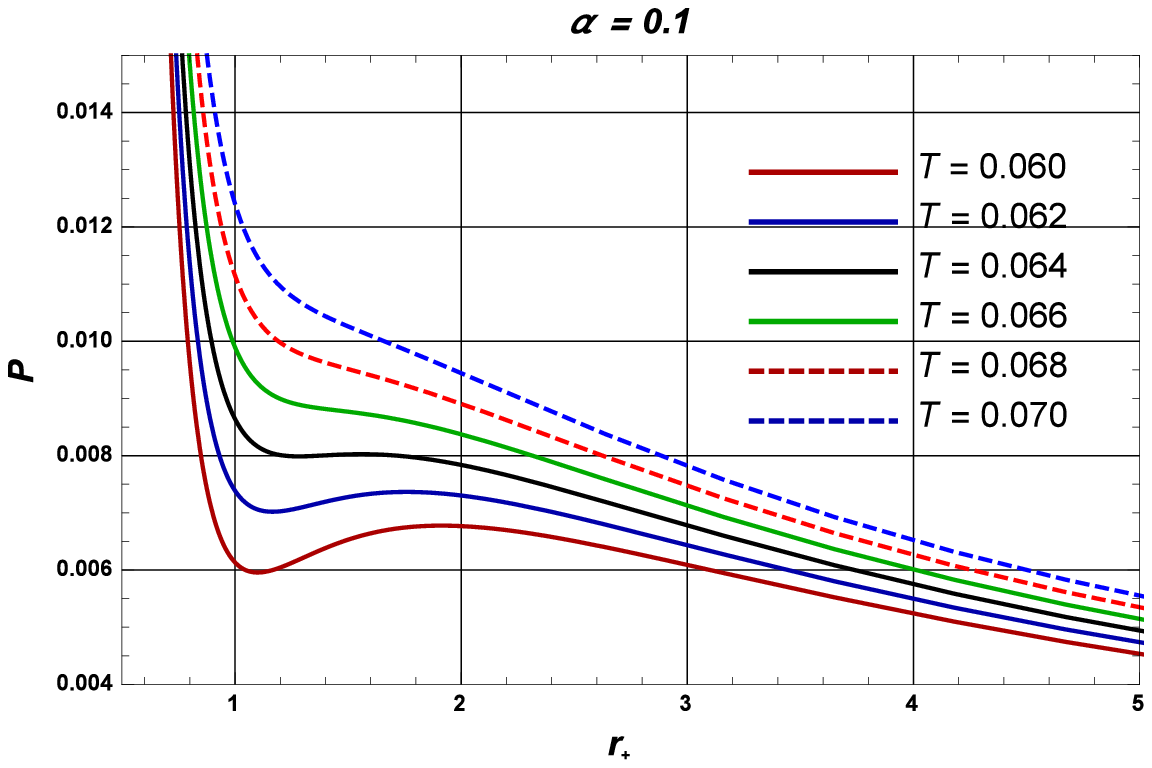}
		\includegraphics[width=.450\linewidth]{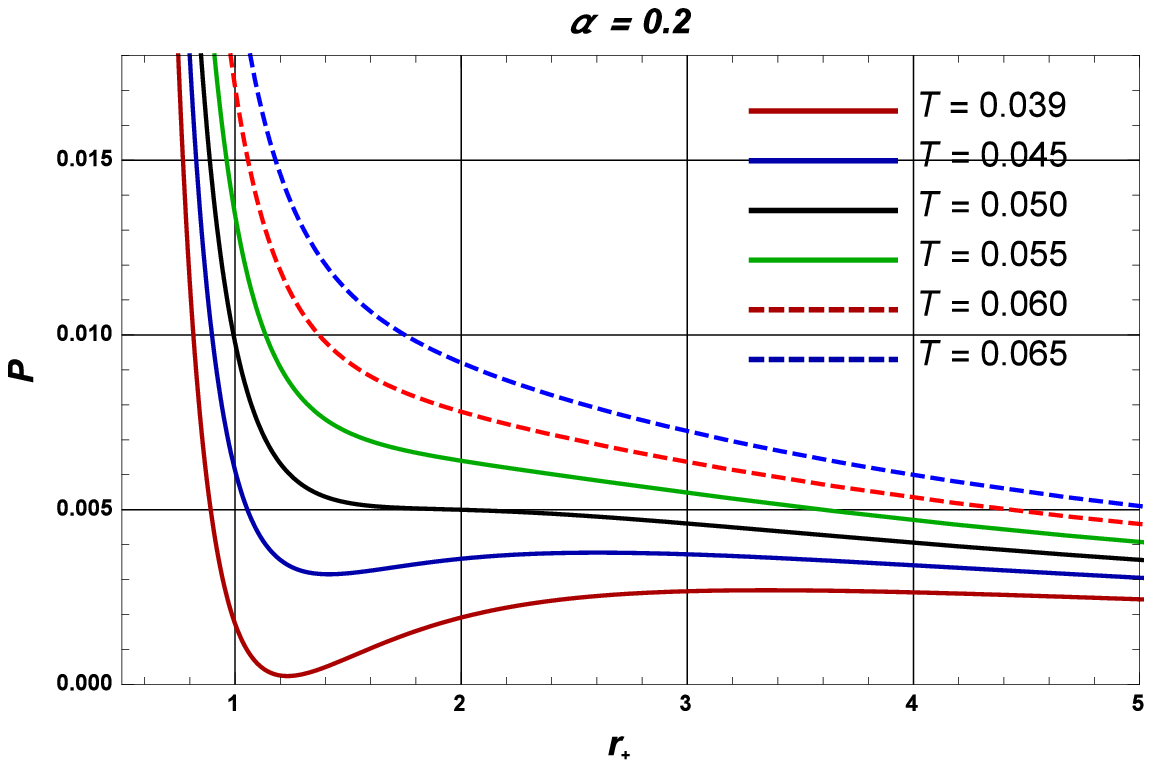} \\
\includegraphics[width=.450\linewidth]{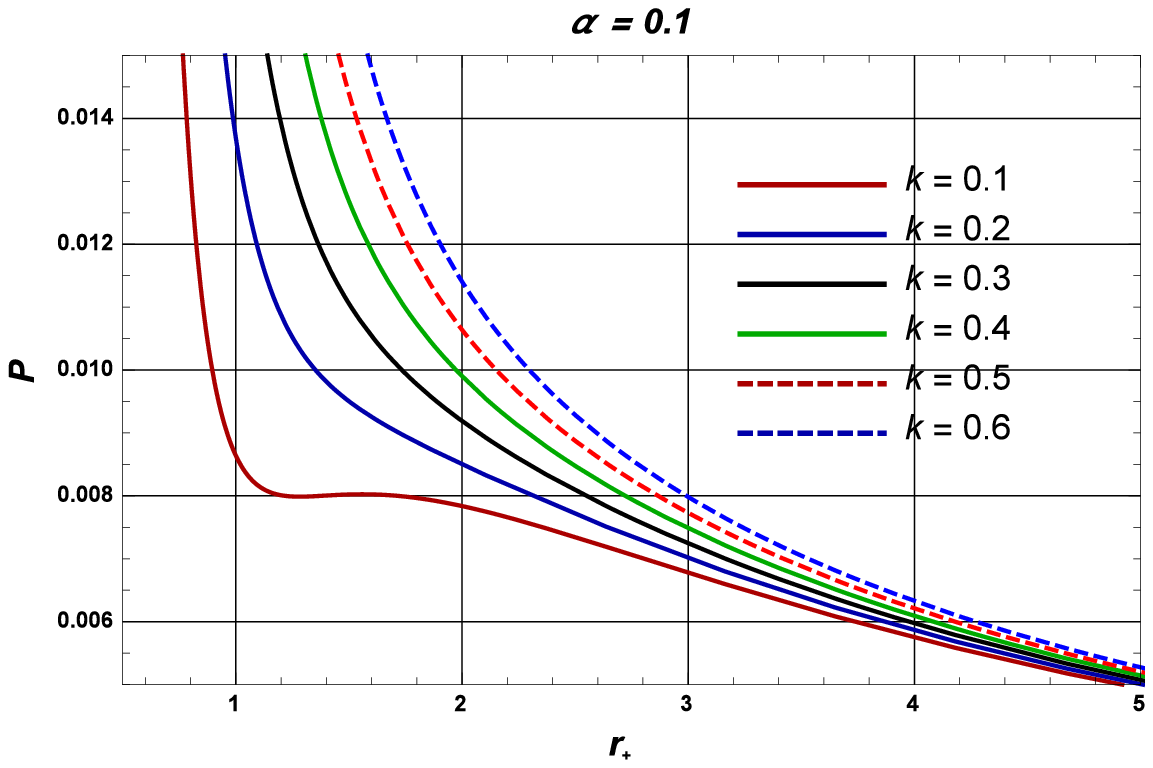}
		\includegraphics[width=.450\linewidth]{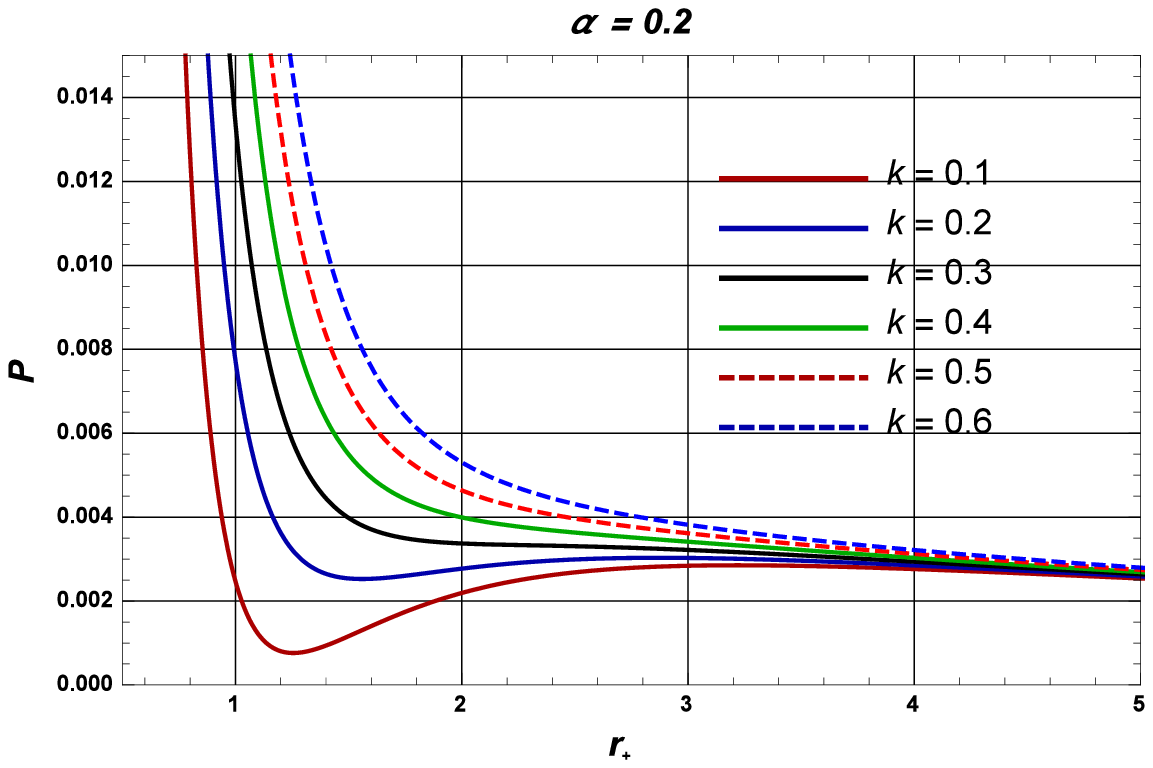} 
	\end{tabular}
	\caption{The plot of pressure versus horizon radius for different values of $T_+$ and $k$ with $\alpha=0.1$ and $\alpha=0.2$ respectively.}
	\label{fig8} 
\end{figure}

\begin{table}[ht]
 \begin{center}
 \begin{tabular}{ l | l   | l   | l   | l   }
\hline
            \hline
  \multicolumn{1}{c|}{ $g$} &\multicolumn{1}{c}{$r_C$}  &\multicolumn{1}{|c|}{$T_C$}  &\multicolumn{1}{c|}{$P_C$} &\multicolumn{1}{c}{${P_C\,r_C}/{T_C}$}\\
            \hline
            \,\,\,\,\,0.1~~ &~~0.8636~~  & ~~0.0884~~ & ~~0.0153~~ & ~~0.1494 \\            
            \,\,\,\,\,0.2~~ &~~1.1144~~ & ~~0.0609~~ & ~~0.0108~~ &  ~~0.1976   \\
            \,\,\,\,\,0.3~~ &~~1.4071~~  & ~~0.0373~~ & ~~0.0075~~ &  ~~0.2829  \\
           \,\,\,\,\,0.4~~ &~~1.717~~ & ~~0.0195~~ & ~~0.0054~~ &    ~~0.4754  \\
            \,\,\,\,\,0.5~~ &~~2.039~~  & ~~0.0063~~ & ~~0.0040~~ &    ~~1.2946   \\
\hline
  \multicolumn{1}{c|}{ $k$} &\multicolumn{1}{c}{$r_C$}  &\multicolumn{1}{|c|}{$T_C$}  &\multicolumn{1}{c|}{$P_C$} &\multicolumn{1}{c}{${P_C\,r_C}/{T_C}$}\\
            \hline
 \,\,\,\,\,0.1~~ &~~0.9527~~  & ~~0.0841~~ & ~~0.0153~~ & ~~0.1732 \\          
            \,\,\,\,\,0.2~~ &~~1.2151~~ & ~~0.0501~~ & ~~0.0096~~ &  ~~0.2328      \\
            \,\,\,\,\,0.3~~ &~~1.4022~~  & ~~0.0342~~ & ~~0.0073~~ &  ~~0.2993    \\
            \,\,\,\,\,0.4~~ &~~1.5798~~ & ~~0.0245~~ & ~~0.0061~~ &  ~~0.3858  \\
            %
            \hline 
\hline  
        \end{tabular}
        \caption{Values of critical temperature $T_C$, critical pressure $P_C$ and universal ration $P_C\,r_C/T_C$ corresponding to different values of $g$ and $k$ for   $ \alpha=0.1$.}
\label{t11}
    \end{center}
\end{table}
In order to see the behavior of pressure, we plot expression of pressure (\ref{pv4}) in FIG. \ref{fig8}.

\section{Connection between phase transition and Shadow radius}\label{s5}
  To compute the geodesics in the spacetime, we first focus to the motion of photon in the the horizon of black hole solution (\ref{sol1}).  The Lagrangian for a photon constrained to move in  the  equatorial plane $(\theta=\pi/2)$ can be expressed as
\begin{equation}
{\cal {L}}=-f(r){d t}^2 +g_{rr}{\dot r}^2+\frac{1}{f(r)}{d \theta^2}+r^2 {d \phi}^2.
\label{lag1}
\end{equation}
 The corresponding Hamiltonian is written as
 \begin{equation}
{\cal{H}}=\frac{1}{2}g^{\mu\nu}p_{\mu}p_{\nu}=\frac{1}{2}\left[-\frac{p_t^2}{f(r)}+{p_r}^2f(r)+\frac{p_{\phi}^2}{r^2}\right].
 \end{equation}
The equations of motion    can be calculated using the Hamiltonian formalism as follows
\begin{equation} 
{\dot t}=\frac{\partial {\cal H}}{\partial p_t}=-\frac{p_t}{f(r)},\qquad\qquad {\dot \phi}=\frac{\partial {\cal H}}{\partial p_{\phi}}=\frac{p_{\phi}}{r^2},\quad\text{and}\quad {\dot r}=\frac{\partial {\cal H}}{\partial p_r}={p_r}f(r).
\end{equation}
where dot denotes the derivative with respect to affine parameter. Using equations of motion  and  conserved quantities, the null geodesics equation is given by
\begin{equation}
    {\dot r}^2+V_{eff}(r)=0, \qquad \text{with} \qquad V_{eff}=f(r)\left(\frac{J^2}{r^2}-\frac{E^2}{f(r)}\right).
\end{equation}
For a circular null geodesics which describes the radius of the photon sphere, the effective potential must  follow
\begin{equation}
V_{eff}=0, \qquad  \text{and}\qquad\frac{\partial V_{eff}}{\partial r}=0.
\label{pot}
\end{equation}
The equation of the photon radius ($r_p$) can be estimated by   Eq. (\ref{pot}). But this  equation  (\ref{pot}) can not be solve analytically, so we solve it numerically and the values are tabulated in the TABLE \ref{pr1}.   
\begin{center}
\begin{table*}[ht]
\begin{center}
\begin{tabular}{|l | l r l r l r l r |r r| }
\hline
\hline
\multicolumn{1}{|c|}{ \it{g} } &\multicolumn{1}{c}{  } &\multicolumn{1}{c}{ 0.1 } & \multicolumn{1}{c}{  }& \multicolumn{1}{c}{0.3} &\multicolumn{1}{c}{ }&\multicolumn{1}{c}{0.5} &\multicolumn{1}{c}{ }   & \multicolumn{1}{c|}{0.7} \\
\hline
\,\,\,\,\,\,\,$r_p$\,&~ \,\,\,\,& 0.12\,\,\,\,\,\,\, &~~\,\,\, &  ~~\,\,0.099\,\, &\,\,\,\,&~~\,\,0.097\,\,&\,\,~~\,\,&~~\,\,0.095\,\,
\\
\,\,\,\,\,\,\,$r_s$\,&~ \,\,\,\,& 0.142\,\,\,\,\,\,\, &~~\,\,\, &  ~~\,\,0.098\,\, &\,\,\,\,&~~\,\,0.097\,\,&\,\,~~\,\,&~~\,\,0.096\,\,
 \\
\hline
\,\,\,\,\,\,$k$&\,\,\,\,&\,\,  0.1\,\,\,\,\,\,\, &~~\,\,\,& ~~\,\,  0.3\,\,&\,\,\,\,&\,\,0.5\,\,& \,\,~~\,\,&~~\,\, 0.7\,\,
\\
\hline
\,\,\,\,\,\,\,$r_p$\, &\,\,~\,\,& 0.12\,\,\,\,\,\,\, &~~\,\,\, &  ~~\,\,0.491\,\, &\,\,\,\,&~~\,\,0.711\,\,&\,\,~~\,\,&~~\,\,0.\,\,\\
\,\,\,\,\,\,\,$r_s$\,&~ \,\,\,\,& 0.142\,\,\,\,\,\,\, &~~\,\,\, &  ~~\,\,0.605\,\, &\,\,\,\,&~~\,\,0.844\,\,&\,\,~~\,\,&~~\,\,0.\,\,\\
\hline
\,\,\,\,\,\,$\alpha$&\,\,\,\,&\,\,  0.1\,\,\,\,\,\,\, &~~\,\,\,& ~~\,\,  0.3\,\,&\,\,\,&\,\,0.5\,\,& \,\,~~\,\,&~~\,\, 0.7\,\,
\\
\hline
\,\,\,\,\,\,\,$r_p$\, &\,\,~\,\,& 0.12\,\,\,\,\,\,\, &~~\,\,\, &  ~~\,\,0.111\,\, &\,\,\,\,&~~\,\,0.108\,\,&\,\,~~\,\,&~~\,\,0.\,\,\\
\,\,\,\,\,\,\,$r_s$\,&~ \,\,\,\,& 0.142\,\,\,\,\,\,\, &~~\,\,\, &  ~~\,\,0.122\,\, &\,\,\,\,&~~\,\,0.116\,\,&\,\,~~\,\,&~~\,\,0.\,\,\\
 \hline
\hline
\end{tabular}
\end{center}
\caption{The numerical values of photon radius and shadow radius corresponding to different values of magnetic charge, deviation  parameter and GB coupling constant.} 
\label{pr1}
\end{table*}
\end{center}
We can obtain the shadow  radius of the black hole. The shadow radius of the black hole  is determined by 
\begin{equation}
r_s=\sqrt{\alpha^2+\beta^2}=\frac{L_p}{E}=\frac{r}{\sqrt{f(r)}}|_{r=r_p}.
\label{sr1}
\end{equation}
Here, we plug  the value of $f(r)$   (\ref{sol1}) into above equation (\ref{sr1}) and we substitute the numerical values of $r_p$  as presented in TABLE \ref{pr1}. From the TABLE \ref{pr1}, we notice that the photon radius increases with the deviation parameter and decreases with the $g$ and $\alpha$.
The  celestial coordinates $\alpha$ and $\beta$ \cite{Kumar:2019pjp,Singh:2017vfr,Ahmed:2022qge,Ahmed:2020dzj} for our black hole solution   are given by
\begin{eqnarray}
  {\bf x}=\lim_{r\to \infty}\left(\frac{rp^{\phi}}{p^t}\right),\qquad\text{and}\qquad
{\bf y}= \lim_{r\to \infty}\left(\frac{rp^{\theta}}{p^t}\right).
\end{eqnarray}
{ 
The shadow of the obtained solution  for different values of magnetic charge $(g)$ and   deviation parameter $(k)$ is depicted in FIG. \ref{fig08}.
\begin{figure}[h]
	\begin{tabular}{c c c c}
\includegraphics[width=.45\linewidth]{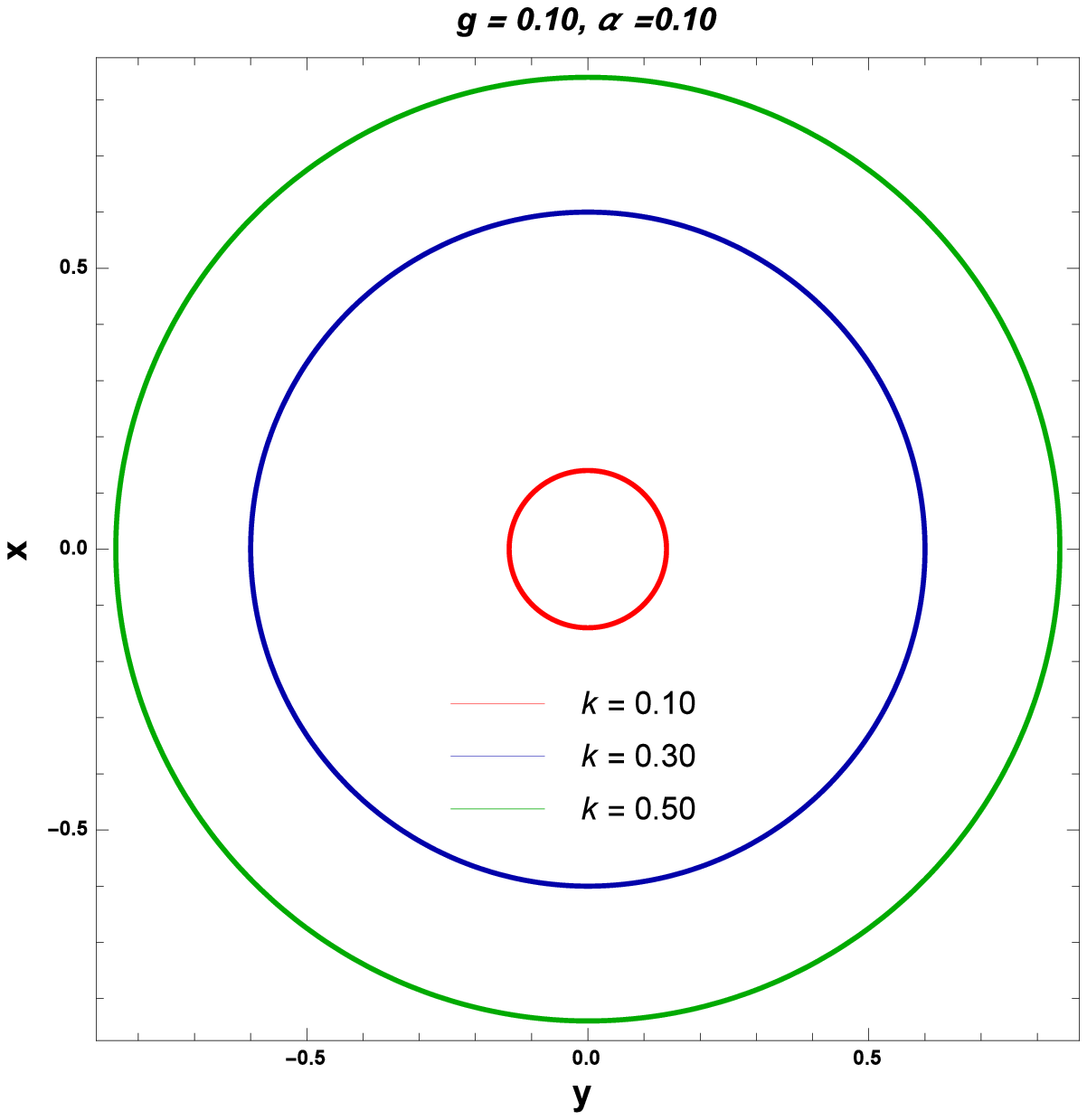}
\includegraphics[width=.45\linewidth]{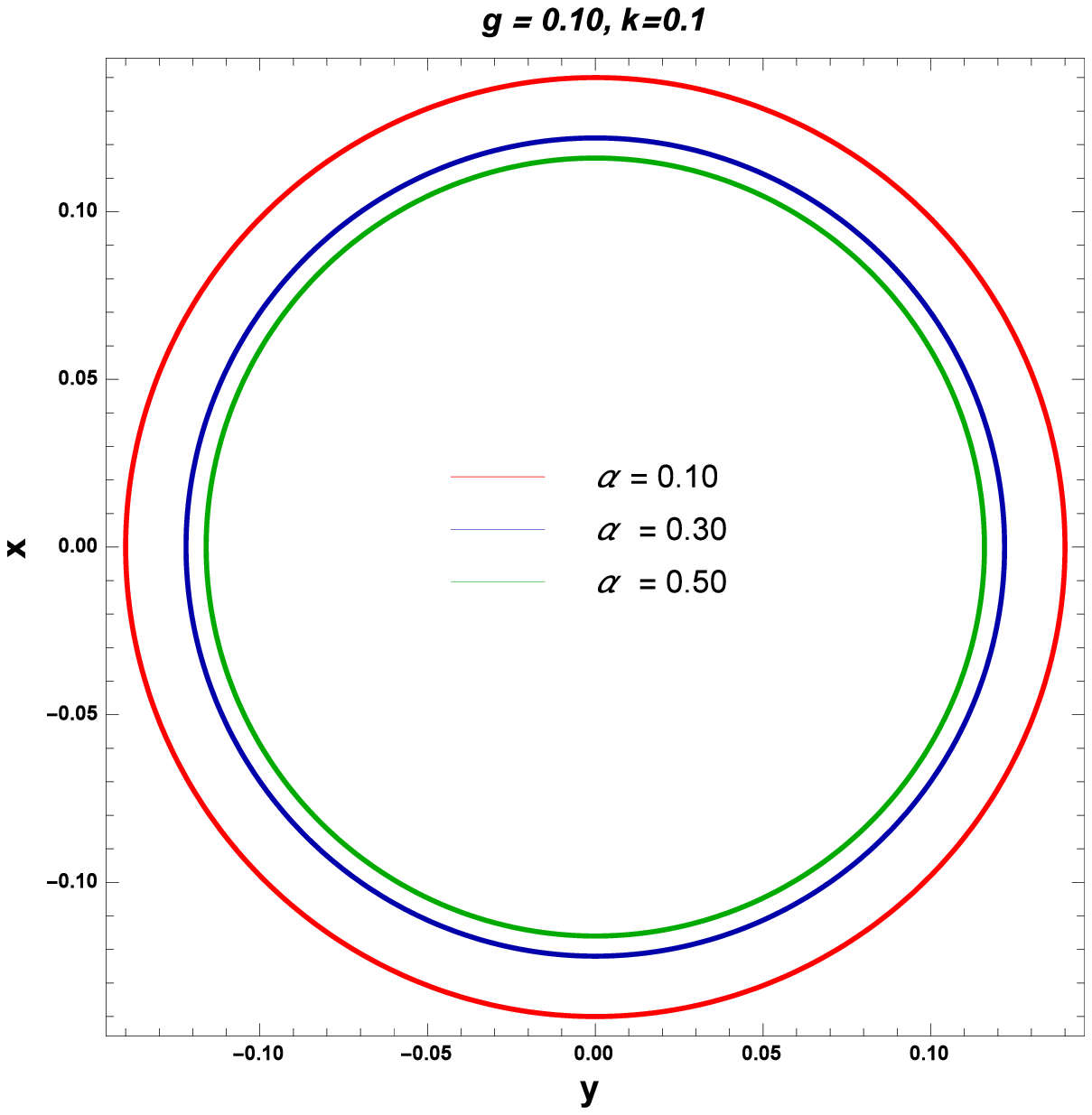} \\
\includegraphics[width=.45\linewidth]{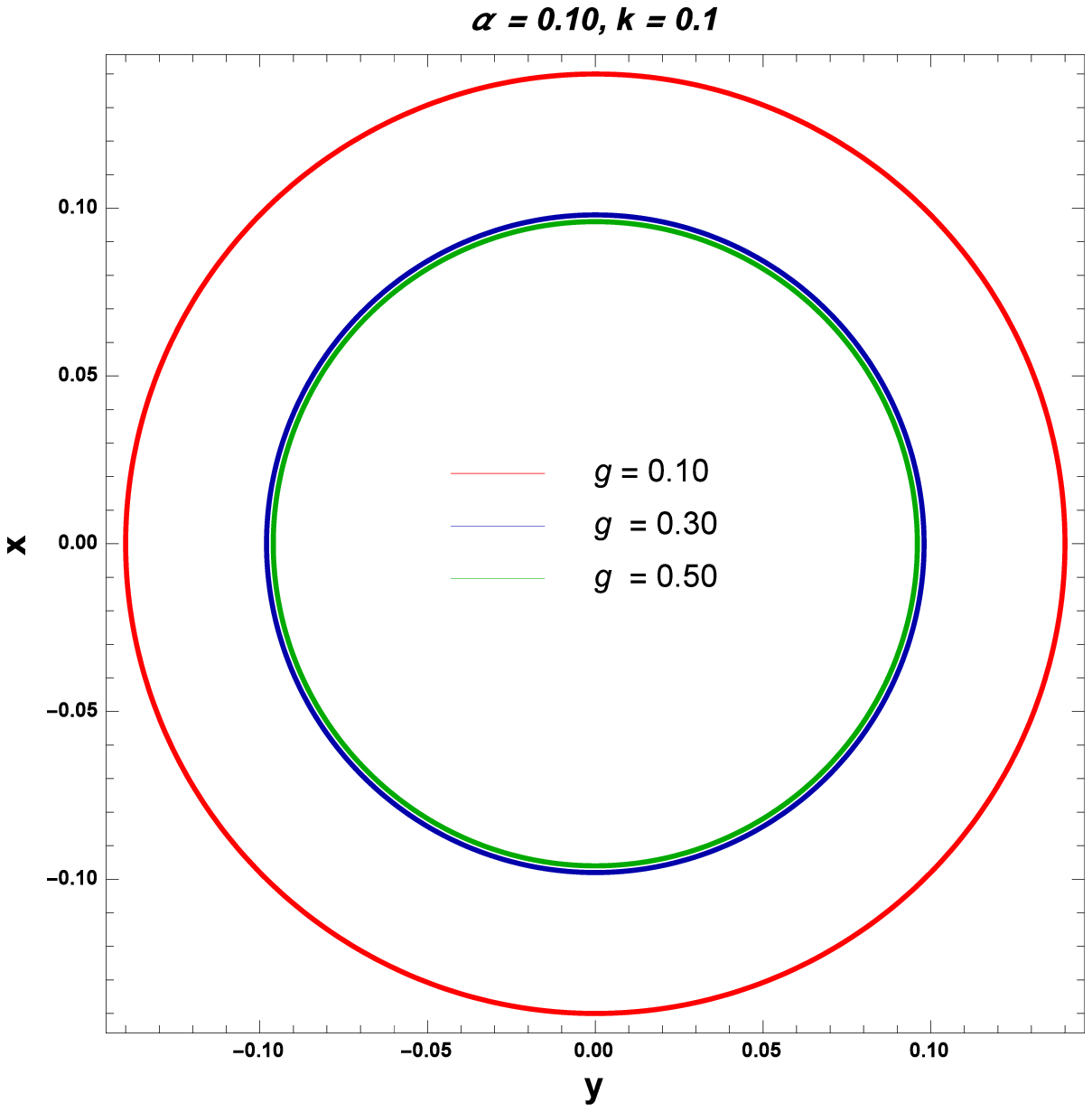}
\includegraphics[width=.45\linewidth]{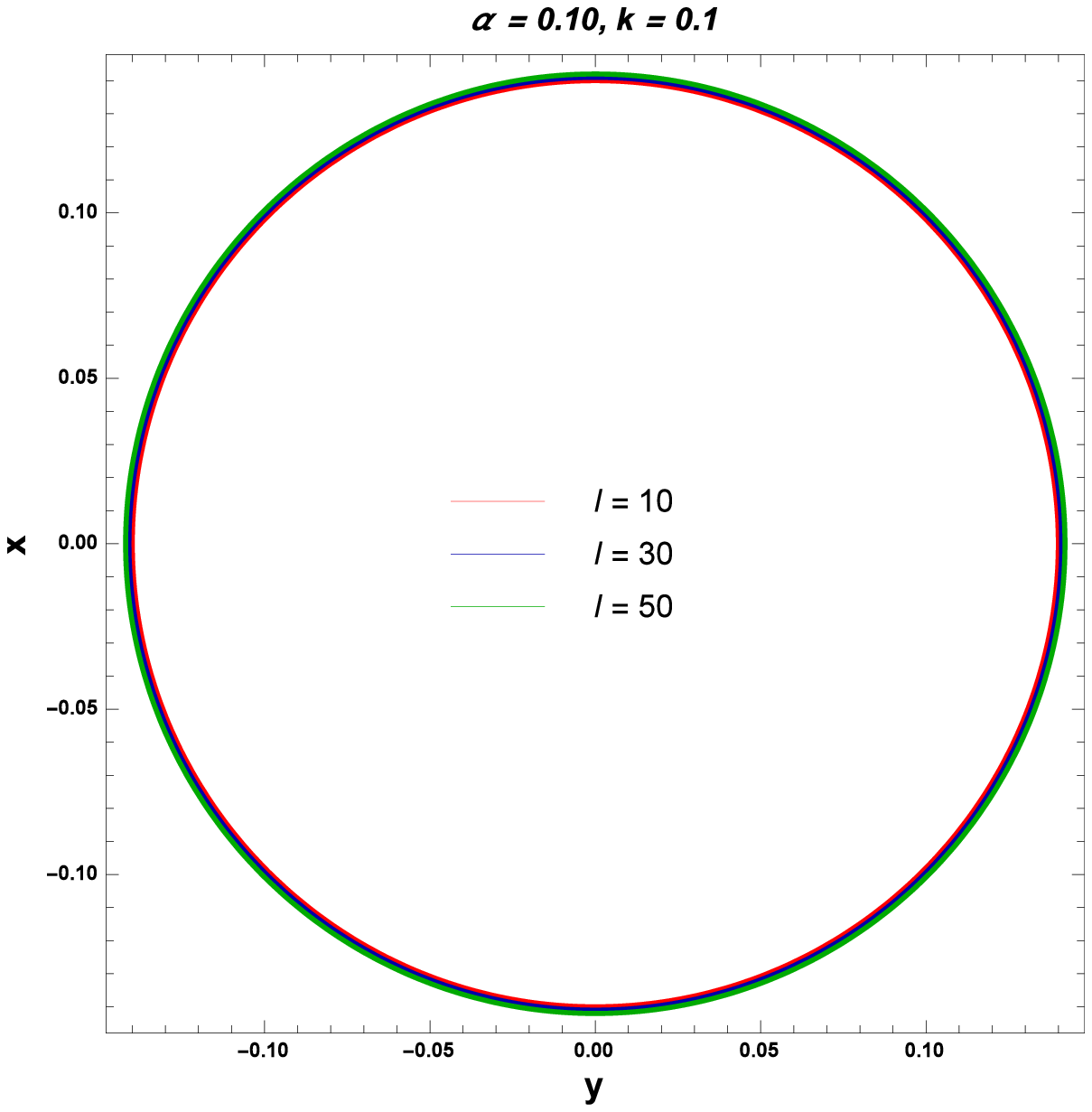} 
	\end{tabular}
	\caption{Plot of shadow with different values of deviation parameter $(k)$, magnetic charge $(g)$, GB coupling $(\alpha)$ and length parameter $l$.}
	\label{fig08} 
\end{figure}
From the figure, we observe the effects of different values of deviation parameter $k$, magnetic charge $g$, GB coupling $\alpha$ and cosmological $(\Lambda=-3/l^2)$ the on the black hole shadow. Here, we find that

\noindent 1)  the size of the shadow increases with increases in the deviation parameter,

\noindent 2)  the size of the shadow decreases with an increase in the magnetic charge and GB coupling constant,

\noindent 3) and the size of the shadow does not change significantly with increase in the value of $AdS$ length.

}

\section{Conclusions}\label{s6}
We have obtained a  general $AdS$ regular black hole solution for EGB gravity coupled with non-linear electrodynamics. Interestingly,  in the limiting cases, our solution 
coincides with AdS EGB black hole, Bardeen black hole, AdS regular black
hole  and AdS Schwarzschild black hole. 
The horizons of this black hole solution are estimated numerically.  We have also analysed the  effects of  deviation parameter, magnetic charge and GB parameter on the horizon structure. The nature of singularities of obtained solution is also checked by calculating  curvature invariants which in turn signifies regular space-time.

Thermodynamics of this general $AdS$ regular black hole solution is also investigated. 
First of all, we have computed total mass of the system and found that (a) switching off deviation parameter the calculated mass of our system coincides with the mass of the $AdS$ EGB Bardeen  black hole \cite{Singh:2020xju}, (b) switching off magnetic charge this coincides with  the mass of  $AdS$ EGB regular  black hole \cite{Singh:2020mty},  (c) switching off both  the deviation parameter and magnetic charge  this identifies  to the mass of $4D$ EGB black hole \cite{gla}, (d) switching off both the GB parameter and deviation parameter  this  reduces to the mass of Bardeen black hole \cite{tzi}, (e)  switching off both the GB parameter and magnetic charge this reduces to  the mass of $AdS$ regular black hole and finally (f)  switching all three GB parameter, deviation parameter and magnetic charge  this identifies with   the mass of $AdS$ Schwarzschild black hole. Furthermore, we have determined Hawking temperature and entropy  of the system. The behavior of temperature is also depicted in plots.  Moreover, we have studied stability of the system by estimating heat capacity.  From the  plot, we have observed that heat capacity has  discontinuity at a   critical radius  which signifies a second-order phase transition.
At critical radius,  the temperature gains its maximum value. This signifies that a phase transition occurs  when the size   gets bigger. 

In addition,  $P-v$  criticality of the obtained black hole solution is also 
discussed. Here, in summery, we found   that, even though  the  critical radius is an increasing function of magnetic charge and   decreasing function of deviation parameter, the universal ratio $\frac{ P_Cr_C}{T_C}$  increases along with both the magnetic charge $g$ and deviation parameter $k$. 

Ultimately, in order to discuss black hole shadow, we have estimated the photon radius and shadow radius and their dependence on  the different values of magnetic charge  and   deviation parameter. To study the behavior of black hole shadow,  we have plotted diagrams as well.

\begin{acknowledgements} 
 One of us (D.V.S.) acknowledges University Grant Commission for the start-up grant (No.30-600/2021(BSR)/1630). 
\end{acknowledgements}   

\section*{Data availability statement} Data sharing not applicable to this article as no datasets were generated or analysed during the current study.
 

\end{document}